\documentclass[prc,aps,12pt,final,notitlepage,oneside,onecolumn,
nobibnotes,nofootinbib,superscriptaddress,showpacs]{revtex4}
\usepackage{amssymb,amsmath}
\usepackage{mathtext}
\usepackage[cp1251]{inputenc}
\usepackage[T2A]{fontenc}
\usepackage[english]{babel}
\usepackage{bm}
\usepackage{array}
\usepackage{graphicx}
\usepackage{calc}
\usepackage{ifthen}
\usepackage{epsfig}

\def\baselinestretch{1.5}

\def\est{e_n}
\def\xip{r_{\parallel}}
\def\dxip{\dot{r}_{\parallel}}
\def\K{\hbox{\bf K}}
\def\E{\hbox{\bf E}}
\def\sn{{\rm sn}}
\def\TA{T_{\rm A}}
\def\X{{\rm X}}
\def\M{{\rm M}}
\def\I{{\rm I}}
\def\Mt{{\rm M}_\perp}
\def\MA{{\rm M}_{\perp A}}
\def\trM{{\rm tr}\,\Mt}

\def\trMA{{\rm tr}\,\MA}
\def\trMRL{{\rm tr}\,{\rm M}_{\perp R,L}}
\def\trMR{{\rm tr}\,{\rm M}_{\perp R}}
\def\trML{{\rm tr}\,{\rm M}_{\perp L}}
\def\be{\begin{equation}}
\def\ee{\end{equation}}
\def\bea{\begin{eqnarray}}
\def\eea{\end{eqnarray}}
\def\eq#1{(\ref{#1})}
\def\bs{\bigskip}
\def\ms{\medskip}
\def\fig#1{Fig.~\ref{#1}}
\def\sec#1{Sec.~\ref{#1}}
\def\tab#1{Tab.~\ref{#1}}

\def\expeps{\epsilon}
\def\siml{\,\hbox{\kern.1em \lower.6ex \hbox{$\sim$} \kern-1.12em
          \raise.6ex \hbox{$<$} }}
\def\simg{\,\hbox{\kern.1em \lower.6ex \hbox{$\sim$} \kern-1.12em
          \raise.6ex \hbox{$>$} }}

\def\Pl{P}
\def\Ql{Q}
\def\bgrk#1{\mbox{{\boldmath $#1$ \unboldmath}}\!\!}

\newcommand{\Table}[4]{
\begin{table}[H]\begin{center}{#3}
\parbox{#2cm}{
\vspace{0.5cm}
\caption[table]{\renewcommand{\baselinestretch}{0.8} \small
                                           \hspace{-0.3truecm}#4}
\label{#1}}
\end{center}
\end{table}
}

\def\Re{{\rm Re}}

\def\R{R}
\def\L{L}

\def\doty{\dot{y}}

\def\dotx{\dot{x}}

\def\l{\label}
\def\A{A}

\def\expeps{\epsilon}
\def\d{{\rm d}}

\begin{document}

\title{Analytic approach to bifurcation cascades\\
       in a class of generalized H\'enon-Heiles potentials}

\author{Sergey N. Fedotkin}
\author{Alexander G. Magner}
\email[]{magner@kinr.kiev.ua}
       
\affiliation{Institute for Nuclear Research, 252028 Prospekt Nauki 47,
             Kiev-28, Ukraine}
           
\author{Matthias Brack}

     \affiliation{Institut f\"ur Theoretische Physik, 
 Universit\"at Regensburg,
             D-93040 Regensburg, Germany}  

\date{\today}

\begin{abstract}
We investigate the bifurcation cascades of a linear librational
orbit in a generalized class of H\'enon-Heiles potentials. The 
stability traces of the new orbits born at its bifurcations are 
found numerically to intersect linearly at the saddle energy 
($e=1$), forming what we term the ``H\'enon-Heiles fans''. In 
the limit close to the saddle energy ($e\to1$), where the 
dynamics is nearly chaotic, we derive analytical asymptotic
expressions for the stability traces of both types of orbits 
and confirm the numerically determined properties of the
generalized ``H\'enon-Heiles fans''. As a bonus of our results, we
obtain analytical approximations for the bifurcation energies
$e_n$ which become asymptotically exact for $e_n\to1$.
\end{abstract}


\maketitle

\section{Introduction}

The approximation of the exact density of states of a quantum 
system in terms of classical periodic orbits via semiclassical 
trace formulae is a fascinating subject which has triggered a 
lot of research (see \cite{gubu,book} and the literature quoted 
therein). It presents a nice illustration of the correspondence
between classical and quantum mechanics, besides allowing one
to approximately determine quantum shell structures in terms of
classical mechanics (see \cite{book} for applications in various
fields of physics). In Hamiltonian systems which are classically
neither regular nor purely chaotic, this semiclassical theory
is enriched -- but also complicated -- by the many facets of 
non-linear dynamics. One of them is the bifurcation of periodic 
orbits when they undergo changes of stability \cite{ozob}.

An essential ingredient to determine the stability of a periodic 
orbit is its so-called stability matrix $\Mt$, appearing in the 
amplitudes of Gutzwiller's trace formula \cite{gutz}, which is 
determined from the linearized equations of motion around the 
periodic orbit. The analytical calculation of $\Mt$ for 
non-integrable systems with mixed dynamics is in general not 
possible; the only non-trivial example is, to our knowledge, 
that of a two-dimensional quartic oscillator \cite{yosh}.

In this paper we investigate the stability matrix $\Mt$ of the 
simplest orbit in a class of two-dimensional potentials
which are a generalization of the famous H\'enon-Heiles (HH)
potential \cite{hhpa} that has become a text-book example of a 
system with mixed classical dynamics.
For small energies the motion is dominated by a harmonic-oscillator 
part and is quasi-regular; at energies close to and above the 
saddles ($e=1$), over which a particle can escape, the motion is 
quasi-chaotic (see, e.g., \cite{gubu,book,hhpa,ford} and the 
literature quoted therein). At all energies below the saddle,
there exists a straight-line librating orbit A oscillating towards
one of the saddles. This orbit undergoes an infinite sequence
of stability oscillations and hence a cascade of bifurcations, 
which can be understood as the main mechanism of the transition 
from regular motion to chaos \cite{davb,mbgu,lamp}. The stability 
traces of the new orbits R and L generated at the bifurcations are 
found numerically \cite{mbgu} to intersect linearly at the saddle 
energy ($e=1$), forming what was termed the 
``H\'enon-Heiles fans''
\cite{bkwf}.

In the present paper we present analytical calculations of the
stability traces of both the A orbit and the new orbits R and L
bifurcating from it. The results are obtained in the limit close
to the saddle ($e\to 1$) and hence asymptotically valid as the
bifurcation energies $e_n$ approach the saddle energy $e=1$.
They confirm analytically the numerical properties of the 
``H\'enon-Heiles fans'' also in the generalized HH potential. 
As a bonus, we obtain analytical expressions for the bifurcation 
energies $e_n$, which are mathematically valid asymptotically for 
$e_n\to 1$, i.e., for $n\to\infty$, and practically for $n\geq 7$ 
within 5 digits. 

In \sec{sechh} we present the generalized H\'enon-Heiles
system and discuss its shortest orbits, the bifurcation cascade 
of the linear A orbit and, in particular, the properties of the
 ``H\'enon-Heiles fans''. In \sec{secstab} we present the
basic ideas of our analytical approach and the essential
results, while the technical details of our calculations
are given in the Appendices \ref{app1} and \ref{app2}.
In \sec{secpert} we present an alternative perturbative
approach for evaluating the stability traces, with the
details given in Appendix \ref{appc2}, and compare
its results with those of the non-perturbative calculations.

\section{Bifurcation cascades in the H\'enon-Heiles system}
\label{sechh}

\subsection{The generalized H\'enon-Heiles Hamiltonian}

In this paper we investigate the following family of Hamiltonians:
\be
H_{GHH} = \frac12\,(p_x^2+p_y^2)+\frac12\,(x^2+y^2)
          + \alpha\left[-\frac13\,y^3+\gamma\,x^2\,y\right]\!,   
\label{ghh}
\ee
where $\gamma\geq 0$ is a parameter specifying specific
members of the family, and $\alpha > 0$ is a chaoticity
parameter that can be scaled away with the energy as shown below. 
For $\gamma=1$, the Hamiltonian \eq{ghh} reduces to the standard 
H\'enon-Heiles (HH) Hamiltonian \cite{hhpa}; we therefore call 
\eq{ghh} here the ``generalized H\'enon-Heiles'' (GHH) Hamiltonian. 
The HH system with $\gamma=1$ has C$_{3v}$ symmetry: it is invariant
under rotations around the origin by $2\pi/3$ and $4\pi/3$, 
and under reflections at three symmetry lines with the angles
$\pm\pi/6$ and $\pi/2$ with respect to the $x$ axis. It exhibits 
three saddles at energy $E_{sad}=1/6\alpha^2$, the equipotential 
lines at $E=E_{sad}$ forming an equilateral triangle. For 
$\gamma\neq1$, the C$_{3v}$ symmetry is lost and only the
reflection symmetry at the $y$ axis remains; there are, however,
still three saddles over which the particle can escape. For 
$\gamma=0$ the system becomes separable and has only one
saddle on the $y$ axis (cf.\ \cite{jkmb,maf}).

After multiplying the Hamiltonian \eq{ghh} by a factor $6\alpha^2$ 
and introducing the scaled variables $x',y',e$ by
\be
x' = \alpha x\,, \qquad y'=\alpha y\,,\qquad 
e = 6\alpha^2 E = E/E_{sad}\,,
\label{scal}
\ee
the scaled Hamiltonian becomes independent of $\alpha$, and for
a given $\gamma$ there is only one parameter $e$ that regulates
the classical dynamics. For simplicity of notation, we omit in the 
following the primes of the scaled coordinates $x,y$ but keep 
using the scaled energy $e$.

For $\gamma=1$, the three saddles are at the scaled energy $e=1$; one
of them is positioned at $x=0$, $y=1$. For $\gamma\neq 1$, the saddle
with energy $e=1$ persists at the same position, while the two
other saddles lie at different energies and are positioned 
symmetrically to the $y$ axis. For a more detailed description of
the topology of the potential \eq{ghh} (and an even larger class
of generalized HH potentials) and its shortest periodic orbits, 
we refer to a forthcoming publication \cite{bttb}.

The shortest periodic orbits of the standard HH system ($\gamma=1$) have 
been extensively discussed in the literature \cite{chur,davb,mbgu,lamp}, 
and their use in semiclassical trace formulae for the quantum 
density of state of the HH system was investigated in 
\cite{bblm,hhun,jkmb,jkwb}.

\subsection{The motion along the A orbit}
\label{secaorbit}

As mentioned above, we use henceforth the symbols 
$x,y$ for the scaled coordinates (corresponding to
$\alpha=1$), along with the scaled energy $e$ given in \eq{scal}.
The equations of motion for the Hamiltonian \eq{ghh} are then
\bea
\ddot{x}(t)+ [1+2\gamma\, y(t)]\,x(t)    & = & 0\,, \label{motionx}\\
\ddot{y}(t)+ y(t)-y^2(t)+\gamma\, x^2(t) & = & 0\,. \label{motiony}
\eea
In the present work we focus on the linear orbit that librates along
the $y$ axis, here called the A orbit. It goes through the origin
$(x,y)=(0,0)$ and towards the saddle at $(x,y)=(0,1)$ which it,
however, only reaches asymptotically for $e\to 1$ with a period
$T_A \to \infty$. Since this orbit has $x_A(t)={\dot x}_A(t)=0$ at all 
times $t$, its equation of motion is
\begin{equation}
\ddot{y}_A(t)+ y_A(t)-y_A^2(t) = 0\,,
\label{motyA}
\end{equation}
which can be solved analytically \cite{jkmb}. We give here the
result in the most general form, relevant for our following 
development, where the initial point along the $y$ axis is
given as $y_0=y_A(t=0)$. The solution is then: 
\begin{equation}
 y_A(t)=y_1+(y_2-y_1)\, \sn^2(z,\kappa)\,,
\label{yAt}
\end{equation}
\begin{equation}
     z = a_\kappa t +F(\varphi,\kappa)\,.
\label{z}
\end{equation}
Here $\sn(z,\kappa)$ is a Jacobi elliptic function \cite{grry} 
with argument $z$; its modulus $\kappa$ and the constant $a_\kappa$ 
are given by
\be
  \kappa = \sqrt{\frac{y_2-y_1}{y_3-y_1}}\,,\qquad
a_\kappa = \sqrt{\left(y_3-y_1\right)\!/6}\,,
\label{qmod}
\ee
in terms of the three real solutions of the equation 
$e=3\,y^2-2\,y^3 \leq 1$ given by
\be
y_1 = 1/2-\cos(\pi/3-\phi/3)\,,\quad 
y_2 = 1/2-\cos(\pi/3+\phi/3)\,,\quad
y_3 = 1/2+\cos(\phi/3)\,, 
\label{yy}
\ee
with $\cos\phi = 1 - 2\,e$. The function $F(\varphi,\kappa)$ in 
\eq{yAt} is the incomplete elliptic integral of first kind with 
modulus $\kappa$, the argument $\varphi$ being determined by the 
initial condition:
\begin{equation}
\varphi=\arcsin\sqrt{\frac{y_0-y_1}{y_2-y_1}}\,.
\label{fi-k}
\end{equation}
$y_1$ and $y_2$ are the lower and upper turning points, respectively, 
of the A orbit along the $y$ axis. The period and the action of the 
(primitive) A orbit are given by
\be
T_A(e) = \frac{2}{a_\kappa}\,\K(\kappa)\,,\qquad
S_A(e) = \frac{12\,a_\kappa}{5\,\alpha^2}\,[\E(\kappa)
         +c_\kappa\K(\kappa)]\,,              
\label{tasa}
\ee
with $c_\kappa = -2(y_3-y_2)(2\,y_3-y_2-y_1)/9$, in terms of the complete 
elliptic integrals of first and second kind, $\K(\kappa)$ and $\E(\kappa)$ 
(we use the notation of \cite{grry}).

Note that in the limit $e\to 1$, we have $y_2\to 1$, $y_3\to 1$ and 
$\kappa\to 1$, so that $\K(\kappa)$ and $T_A$ diverge (while $S_A$ 
remains finite). The A orbit then is no longer periodic (and may be 
called a ``homoclinic orbit'' \cite{ozob}). Expanding $T_A$ around 
$e=1$, one finds the asymptotic form \cite{mbgu}
\be
T_A(e) \; \approx \; \widetilde{T}_A(e) = 
{\rm ln}\left(\frac{432}{1-e}\right). 
                                          \qquad (e\to 1)
\label{Tas}
\ee

\subsection{The bifurcation cascade of the A orbit in the standard
            HH potential}

While approaching the saddle as $e\to 1$, the A orbit undergoes an
infinite cascade of pitchfork bifurcations, giving birth to a
sequence of new orbits R$_5$, L$_6$, R$_7$, L$_8$, \dots. This
scenario, which has some similarities to the Feigenbaum scenario
\cite{feig}, was discussed extensively in \cite{mbgu}, and the
analytical forms of the newborn R and L orbits in terms of periodic
Lam\'e functions were discussed in \cite{lamp}. 

In \fig{zoom} we show the traces of the stability matrix $\Mt$,
defined in \eq{M1} below, of the A orbit and the orbits bifurcated
from it, plotted versus energy $e$. Whenever 
$\trM=2$, a bifurcation occurs. We see the successive bifurcations 
at increasing energies $e_n$; upon repeated zooming the upper end of 
the energy scale near $e=1$ (from bottom to top), the pattern repeats 
itself in a self-similar manner. The bifurcation energies $e_n$ form
a geometrically progressing series (see \cite{mbgu,lamp} for details). 
cumulating at the saddle energy ($e=1$) such that
$e_5({\rm R}_5) < e_6({\rm L}_6) < e_7({\rm R}_7) < \dots < 1$,
where the parentheses contain 
the names of the new orbits born at the pitchfork bifurcations. These 
are alternatively of R type (rotations) and of L type (librations). 
(The subscripts in the orbit names indicate the Maslov indices 
appearing in the semiclassical trace formulae; the index of 
the A orbit increases by one unit at each bifurcation.)
Due to the discrete symmetries of the system, all these pitchfork
bifurcations are isochronous and hence not generic (cf.\ \cite{bttb}).

In \fig{hhtrmt}, we show again $\trM$ -- in the following briefly
termed the ``stability traces'' -- of the same orbits, but
this time plotted versus their respective periods $T$. On this
scale, $\trMA(\TA)$ (shown by the heavy line) is numerically 
found \cite{davb} for large $\TA$ to go like a sine function; 
its period $\Delta T=3.6276$ was shown in \cite{mbgu} to be
given analytically by $\Delta T = 2\pi/\!\sqrt{3}$. The exact
calculation of the function $\trMA(\TA)$ is, however, not trivial
at all. It is one of the objects of our present investigations
(see Sec.\ \ref{sectrma}).

\subsection{The ``H\'enon-Heiles fans''}

An interesting property of the stability traces of the R and L orbits
born at the bifurcations, which has been observed numerically
\cite{mbgu} and termed the ``H\'enon-Heiles fan'' structure
\cite{bkwf}, is emphasized in \fig{hhfane}. Here we plot the stability
traces of the primitive A orbit and the first three primitive pairs of
R and L orbits versus the scaled energy $e$. We note two prominent
features (which can also be recognized in \fig{zoom}):\\ 
($i$) The functions $\trMRL(e)$ are approximately linear up to 
(and even beyond) the barrier energy $e=1$.\\
($ii$) The curves $\trMRL(e)$ intersect at $e=1$ in one point 
each for all R and L type orbits with Maslov indices $>8$, positioned 
at the values $2\pm d$ with
$d=6.183\pm 0.001$, thus forming two fans emanating from these points. 
The uncertainty in the parameter $d$ comes from the numerical difficulty 
of finding periodic orbits (which was done using a Newton-Raphson 
iteration procedure) close to bifurcations; our result for $d$ was 
obtained for R$_n$ and L$_{n'}$ orbits with $9\leq n,n'\leq 13$,
evaluated at $e=1$. The upper limit $n=13$ is due to the numerical
problems only; we expect that the same value $d=6.183\pm
0.001$ holds also for all higher $n$.

We found exactly the same types of ``HH fans'' for the generalized
HH systems given by the Hamiltonian \eq{ghh} for the bifurcation
cascade of the A orbit along the $y$ axis, whereby the slopes
of the fans and hence the value of $d$ depend on the parameter
$\gamma$. The ``GHH fans'' can be described, for large enough $n$, 
by the empirical formula
\be
\trMRL^{(emp)}(e) \; = \; 2 \mp c_{RL}(\gamma)\, \frac{(e-e_n)}{(1-e_n)}\,,
                       \qquad (e\geq e_n)  
\label{trMTRLnum}
\ee
where the negative and positive sign belongs to the R and L type orbits, 
respectively. At $e=1$ the curves $\trMRL(e)$ intersect linearly at
the two values $\trMRL(1)=2\mp c_{RL}(\gamma)$, so that the parameter
$d$ given above for the standard HH potential is $d=c_{RL}(1)$. The 
numerical values for $c_{RL}(\gamma)$ are shown by crosses in 
\fig{fanslope} below.

The main goal of our paper is to find analytical support for these
numerical findings. In Sec.\ \ref{secstab} we will, indeed, confirm 
the empirical formula \eq{trMTRLnum} analytically in the asymptotic
limit $e\to 1$.

\section{Asymptotic evaluation of stability traces}
\label{secstab}

In this section we derive analytic expressions for $\trM(e)$ of
the A, R and L orbits in the GHH system, which are valid in the 
asymptotic limit $e\to 1$, i.e., close to the barrier. Before
presenting them in \sec{sectrma} and \sec{sectrmrl}, we recall
the definitions of the stability matrix $\trM$ and of the
monodromy matrix M of which it is a submatrix.

\subsection{Monodromy and stability matrices}
\label{monstab}

\subsubsection{Stability matrix and the Hill equation}
\label{sechill}

The analytical calculation of the stability matrix $\Mt$ of a
periodic orbit in a non-integrable system is in general a
difficult task. We recall that the stability matrix is obtained
from a linearization of the equations of motion and defined by
\be
\delta\bgrk{\xi}_\perp(T) = \Mt\,\delta\bgrk{\xi}_\perp(0)\,,
\label{M1}
\ee 
where $\delta\bgrk{\xi}_\perp(t)$ is the $(2N-2)$-dimensional phase-space
vector of infinitesimally small variations transverse to the
given periodic orbit ($N$ being the number of independent
degrees of freedom), and $T$ is the period of the orbit. 
For $N=2$ dimensional systems, we may choose $\bgrk{\xi}_\perp(t)=(q,p)$ 
where $q$ is the coordinate and $p$ the canonical momentum 
transverse to the orbit in the plane of its motion. $(q,p)$ then 
form a ``natural'' canonical pair of Poincar\'e variables, normalized 
such that $(q,p)=(0,0)$ is the fixed point of the periodic orbit 
on the projected Poincar\'e surface of section (PSS). For 
two-dimensional Hamiltonians of the form ``kinetic + potential energy'': 
$H=T+V$ (and particles with mass $m=1$, so that $p=\dot q$), the 
Newtonian form of the linearized equation of motion for $q(t)$ 
becomes the Hill equation (see the text book \cite{mawi} for an 
explicit discussion)
\be
{\ddot q}(t) + V_{qq}(t)\,q(t) = 0\,, 
\label{hill}
\ee
where $V_{qq}(t)$ is the second partial derivative of the potential 
$V$ with respect to $q$, taken along the periodic orbit, and the
two-dimensional stability matrix is given by
\be
\left( \begin{array}{c} q(T) \\ \dot{q}(T) \end{array} \right)
= \Mt \left( \begin{array}{c} q(0) \\ \dot{q}(0) \end{array} \right).
\label{M2}
\ee
For isolated periodic orbits, solutions of \eq{hill} with
$q(t)\neq 0$ are in general not periodic. However, when the
orbit undergoes a bifurcation, \eq{hill} has at least one
periodic solution which describes the transverse motion of
the new orbit born at the bifurcation; the criterion for the 
bifurcation to occur is $\trM=+2$ (cf.\ \cite{mawi}).

For particular systems, the Hill equation \eq{hill} may become a 
differential equation with known periodic solutions. For the GHH 
systems under investigation here, the Hill equation for the A orbit
directed along the $y$ axis is given by \eq{motionx}, with $y(t)$
replaced by $y_A(t)$ in \eq{yAt}, and becomes the Lam\'e equation
(see, e.g., \cite{erde})
whose periodic solutions are the periodic Lam\'e functions (see
\cite{lamp} for the details). However, the elements of $\Mt$ in 
\eq{M2} can in general not be found analytically. One of the rare 
exceptions is that of the coupled two-dimensional quartic oscillator 
for which Yoshida \cite{yosh} derived an analytical expression for 
$\trM$ as a function of the chaoticity parameter (cf.\ \cite{bfmm}).

Magnus and Winkler \cite{mawi} have given an iteration scheme
for the computation of $\trM$ for periodic orbits in smooth
Hamiltonians. We have tried their method for the A orbit in
the HH system, but we found \cite{mima} that its convergence is 
too slow for computing $\trMA(e)$ with a sufficient accuracy that
would allow to deduce the properties of the HH fans. However,
in the limit $e\to 1$, it is possible to use an asymptotic
expansion of the function $\sn$ appearing in $y_A(t)$ of
\eq{yAt}, which allows us to compute
$\trMA(e)$ analytically, as discussed in \sec{sectrma}.

\subsubsection{Matrizant and monodromy matrix}
\label{secmonod}

For curved periodic orbits -- such as the R and L type orbits
bifurcating from the A orbit in the GHH systems -- which usually can 
only be found numerically, the phase-space variables $\bgrk{\xi}_\perp$ 
transverse to the orbit used in the definition \eq{M1} of the 
stability matrix $\Mt$
cannot be constructed analytically. Instead, one must in general use
Cartesian coordinates and resort to the full monodromy matrix M
defined below. For $N=2$, one first linearizes the equations of motion 
to find the matrizant $\X(t)$ which propagates small perturbations of 
the full phase-space vector $\bgrk{\xi}(t)$ defined by 
\be
\bgrk{\xi}(t) = \{x(t),y(t),\dot{x}(t),\dot{y}(t)\}
\ee
from their initial values at $t=0$ to a finite time $t$:
\be\l{defX}
\delta\bgrk{\xi}(t) = \X(t)\,\delta\bgrk{\xi}(0)\,.
\ee
For a Hamiltonian of the form
$H(x,y,\dot{x},\dot{y})=\frac12\,(\dot{x}^2+\dot{y}^2)+V(x,y)$, the
differential equation for $\X(t)$ is
\be
\frac{\rm d}{{\rm d}t}\, \X(t) \; = \;
\left( \begin{array}{cc} \;\;0 & \I_2 \\
            \!\!-{\rm U}(t) & 0 \end{array}\right)\X(t)
\label{Xoft}
\ee
with the initial conditions
\be
\X(0) = \I_4\,, 
\ee
where $\I_2$, $\I_4$ are the two- and four-dimensional unit matrices and 
U$(t)$ is the two-dimensional Hessian matrix of the potential,
taken along the periodic orbit ($po$):
\be
{\rm U}_{ij}(t) = \frac{\partial^2 V}
                  {\partial x_i \partial x_j}\{x(t),y(t)\}_{po}\,,
                  \qquad (x_i,x_j= x,y)\,.
\ee

Having solved \eq{Xoft}, the monodromy matrix M of the given periodic
orbit with period $T$ is defined by
\be
\M = \X(T)\,.
\label{defM}
\ee
In an autonomous system, M has always two unit eigenvalues
corresponding small initial variations along the periodic orbit
and transverse to the energy shell. After a transformation
to an ``intrinsic'' coordinate system, in which one of the 
coordinates is always in the direction $r_\parallel$ (with 
momentum $p_\parallel=\dxip$) of the periodic orbit \cite{gutz}, 
M can be brought into the form
\begin{equation}
 M = \left( \begin{array}{cc}
        \Mt  & {\bf ...} \\
        {\bf 0} & \left (\begin{array}{cc} 1 & ... \\ 0 & 1 \end{array}\right)
               \end{array} \right),                             \label{Mtil}
\end{equation}
where the dots denote arbitrary non-zero real numbers and $\Mt$ 
is the stability matrix. The diagonal elements in the lower right
block of \eq{Mtil} then correspond to 
\be
\M_{\xip\xip}=\M_{\dxip\dxip}=1\,.
\label{parid}
\ee

The transformation to such an intrinsic coordinate system is 
quite non-trivial \cite{ekwi} and not unique. For curved orbits 
it can in general only be found numerically and is therefore not 
suitable for analytical calculations. For the curved R and L
orbits of our system, we therefore have to resort to the full
monodromy matrix M \eq{defM} via the solution of \eq{Xoft}.
For the evaluation of their stability traces, we only need the 
diagonal elements of M and can then use the obvious relation
$\trM = {\rm tr}\,\M - 2$.

\subsection{Asymptotic evaluation of the stability trace $\trMA(e)$ for $e\to 1$}
\label{sectrma}

In the limit $e\to 1$, where the modulus $\kappa$ defined in
\eq{qmod} goes to unity, we may approximate $y_A(t)$ by 
the leading term in the expansion of the function $\sn(z,\kappa)$
around $\kappa=1$ (see \cite{grry})
\be
\sn(z,\kappa) \approx {\rm tanh}(z)\,. \qquad (\kappa\to 1) 
\label{fact}
\ee
Since the function $\tanh(z)$ is not periodic, we have to approximate 
$y_A(t)$ in two portions. Taking $t_2$ as the time where the orbit 
passes through its maximum at $y_2$, i.e.,
\be
y_A(t_2)=y_2 \quad \Longleftrightarrow \quad t_2=[\K(\kappa)-F(\varphi,\kappa)]/a_\kappa\,,
\ee
we define the asymptotic expression for the A orbit over one period by
\be
{\widetilde y}_A(t)=\Theta(t_2-t)\,Y_1(t)+\Theta(t-t_2)\, Y_2(t)\,, 
                    \qquad 0 \leq t \leq T_A\,, 
\label{yTheta}
\ee
where the functions $Y_1(t)$ and $Y_2(t)$ are given by
\bea
Y_1(t)=y_1 +(y_2-y_1)\, \tanh^2(z)\,, \qquad
Y_2(t)=y_1 +(y_2-y_1)\, \tanh^2(z-2\,\K(\kappa))\,,
\label{six}
\eea
with $z$ given in \eq{z}. Although the function \eq{yTheta} is not 
analytic at $t=t_2$, it suffices to find an asymptotic expression for 
$\trMA(e)$ valid for $e\to 1$. 

The details of our calculation are given in Appendix \ref{app1}.
The analytical asymptotic result for $\trMA\,(e)$ is given in
\eq{trMTA2} in terms of associated Legendre functions.
In the limit $e\to 1$, the energy dependence of $\trMA\,(e)$ goes 
only through the period $\TA(e)$: 
\begin{equation}
\trMA\,(e) \; \approx \; \trMA^{(as)}(e)
                       = \trMA^{(as)}\,(T_A(e),\gamma)\,, \qquad (e\to 1)
\label{trMAeas}
\end{equation}
where $\trMA^{(as)}\,(T_A,\gamma)$ is a universal function given by
\be
\trMA^{(as)}(T_A,\gamma) = +2\,|\widetilde{F}_A(\gamma)| 
            \cos\!\left[\sqrt{1+2\gamma}\,T_A-\widetilde{\Phi}_A(\gamma)\right]\!.
\label{trMTAf}
\ee
The phase function $\widetilde{\Phi}_A(\gamma)$ is defined through
Eqs.\ \eq{FA} and
\eq{fi}, and the amplitude function $|\widetilde{F}_A(\gamma)|$ 
is given explicitly in \eq{omfas}.
We recall that $\gamma$ is the potential parameter of the GHH potential 
\eq{ghh} is $\gamma=1$ for the standard HH potential.
For this case, the result \eq{trMTAf} becomes 
\begin{equation}
\trMA^{(as)}\,(T_A,1) = 2.68043976\,\cos(\!\sqrt{3}\, T_A+1.56782696)\,, 
\label{trMTHH}
\end{equation}
where the numerical constants have been calculated for $\gamma=1$.
The period of the $\cos$ function in \eq{trMTHH} was correctly 
shown in \cite{mbgu} to be $2\pi/\!\sqrt{3}$, but the phase 
$\widetilde{\Phi}_A(\gamma=1)$ and the amplitude $2|\widetilde{F}_A(\gamma=1)|$ 
were only obtained numerically. The asymptotic relation \eq{trMAeas} 
had already been observed numerically in \cite{davb,mbgu}.

The result \eq{trMTHH} is shown in \fig{hhtrmas} by the dotted line 
and compared to the exact numerical result from \cite{mbgu}, shown
by the solid line. We see that the agreement becomes nearly perfect 
for $\TA\simg 10.5$, corresponding to $e\simg e_6$.
The asymptotic result \eq{trMAeas}, \eq{trMTAf} allows us to give
analytical expressions for the bifurcation energies $e_n$ in the
asymptotic limit $e_n\to1$. The pitchfork bifurcations of the A orbit 
occur when $\trMA=+2$. We therefore define approximate bifurcation
energies $e_n^*$ by
\be
\trMA^{(as)}\,(T_A(e_n^*),\gamma) = +2\,.
\label{eqeas}
\ee
Using the asymptotic form of $T_A(e)$ in \eq{Tas} and \eq{trMTAf}, we 
can give the solutions of \eq{eqeas} in the following formulae
\bea
e_{2k-1}^* & \approx & 1 - 432\,\exp\{-[\widetilde{\Phi}_A(\gamma) - \arccos(1/|\widetilde{F}_A(\gamma)|)
                     + 2\pi k]/\!\sqrt{1+2 \gamma}\},
                     \qquad (\hbox{R})\nonumber\\
e_{2k}^*   & \approx & 1 - 432\,\exp\{-[\widetilde{\Phi}_A(\gamma) + \arccos(1/|\widetilde{F}_A(\gamma)|)
                     + 2\pi k]/\!\sqrt{1+2 \gamma}\},
                     \qquad (\hbox{L})
\label{enas}
\eea
where $k=3,4,5,\dots$, and the odd numbers $n=2k-1$ refer to the R type
and even $n=2k$ to the L type bifurcations, respectively. For 
$e_n^*$ sufficiently close to 1, i.e., for large enough $n$ the above
values should reproduce the numerically obtained ``exact'' values $e_n$.

This is demonstrated for $\gamma=1$ in \tab{tabenas}. In the second 
column we give the resulting values of $e_n^*$ with $5 \leq n \leq 16$ 
for the standard HH system, and in the third column we 
reproduce their numerical values $e_n$ obtained in \cite{lamp} 
as roots of the equation $\trMA(e_n)=+2$. As we see, the
asymptotic results $e_n^*$ approach the numerical values $e_n$
very well already starting from $n=7$, as could be expected from
\fig{hhtrmas}. In view of the numerical difficulties in determining 
the $e_n$ from a search of periodic orbits (cf.\ the remarks after 
\fig{hhfane}), the agreement is very satisfactory for all $n\geq 7$.

This is in itself a remarkable result, because we are not aware
of any analytical results for bifurcation energies (or bifurcation
values of any chaoticity parameter) in non-integrable Hamiltonian 
systems, except for the coupled two-dimensional quartic oscillator 
(see \cite{lamp,bfmm}). In the present case, the bifurcation energies
$e_n$ can be related to the eigenvalues of the Lam\'e equation.
These can, in principle, be given by infinite continued fractions
\cite{inc1}, but their determination is hereby only possible
numerically by iteration, which becomes even less accurate than the
numerical solution of $\trMA(e_n)=+2$ as done in
\cite{lamp}. The analytical expressions \eq{enas} therefore
represent an important achievement of this paper.

\subsection{Asymptotic evaluation of $\trMRL(e)$ for $e\to1$}
\label{sectrmrl}

For the stability traces of the R and L orbits we need, as mentioned 
in \sec{secmonod} above, to know the diagonal elements of the 
full monodromy matrix M, i.e., the elements X$_{ii}(t=T)$ with
$i=x,y,\dotx,\doty$. Since the equations \eq{Xoft} couple 
all 16 elements of X$(t)$, this is still a considerable task.
It can, however, be simplified considerably in the asymptotic 
limit $e\to1$. First, we can make use of the ``frozen $y$ motion
approximation'' (in short: ``frozen approximation'', FA) introduced
in Refs.\ \cite{mbgu,lamp}. It exploits the fact that near the
bifurcation energies $e_n$ at which the R and L orbits are born,
their motion in the $y$ direction is close to that of the bifurcating
A orbit and, for increasing energy $e$, changes only very little.
It can be shown (cf.\ \cite{bfmm} and \sec{secpert} below) that 
this may correspond to the first order in a perturbative expansion 
in the parameter $e-e_n$, valid to leading order in the small 
quantity $1-e_n$. Second, we can exploit some
symmetry relations between the elements of M if the initial point
at $t=0$ for the calculation of X$(t)$ is chosen as the upper turning 
point in the direction of the A orbit, i.e., its maximum along the 
$y$ axis. These symmetry relations are derived in \sec{app2suba};
their main consequence is that we only need to calculate the
4$\times$4 submatrix of X$_{ij}(t)$ with spatial indices $i,j=x,y$,
and that we have the asymptotic equality $\trMRL\approx 2$M$_{yy}$
for $e\to1$, see \eq{tracR1}. As shown below, these symmetry relations 
can be used also beyond the FA, and only in order to simplify them
some properties of the FA will be exploited in our further derivations.

With these approximations, the calculation of $\trMRL(e)$ proceeds
similarly as that of $\trMA(e)$ discussed in the previous section;
its details are presented in Appendix \ref{app2}. The analytical
result is given in \eq{trmmyxanas} in terms of associated 
Legendre functions. After their expansion in the asymptotic limit
$e\to1$ we obtain the result
\be
\trMRL^{(as)}(e) \; = \; 2 \mp c_{RL}(\gamma)\,\frac {(e-e_n)}{(1-e_n)}\,, 
                         \qquad (e\geq e_n \to 1)
\label{trMTRLas}
\ee
where the ``$-$'' and ``$+$'' sign belongs to the R and L orbits,
respectively. The slope function $c_{RL}(\gamma)$ is found 
analytically to be
\be
c_{RL}(\gamma) = \frac{4\sqrt{1+2\gamma}}
                 {\sinh[2\pi\sqrt{1+2\gamma}]}
                 \cosh\left(\frac{\pi}{2}\sqrt{48\gamma-1}\right).
\label{crl}
\ee
Eq.\ \eq{trMTRLas} has exactly the functional structure of the 
empirical ``GHH fan'' formula \eq{trMTRLnum}. Mathematically, 
it holds asymptotically in the limit $e_n\to 1$ 
to leading order in the small parameter $\sqrt{1-e_n}$. We 
emphasize that this result confirms also the numerical finding that, 
for large enough $n$ (practically, for $n>8$) the functions 
$\trMRL(e)$ are linear in $e$ from $e_n$ up to at least $e=1$.

In \fig{fanslope} we show by crosses the values of $c_{RL}(\gamma)$, 
evaluated numerically from the stability of the R and L orbits at 
$e=1$, as a function of $\gamma$. The solid line shows the
analytical result \eq{crl}. In the lower part of the figure, we
show the region of small $\gamma$. 
The curve $c_{RL}(\gamma)$ goes through zero with a finite slope 
which can easily be found by Taylor expanding \eq{crl} after the 
replacement $\cosh{(\pi\sqrt{48\gamma-1}/2)} \rightarrow
\cos{(\pi\sqrt{1-48\gamma}/2)}$. The slope at $\gamma=0$ becomes
\begin{equation}
c'_{\!RL}(0) = \left. \frac{\d}{\d\gamma}\,c_{RL}(\gamma)\right|_{\gamma=0}
           = \frac {48\pi}{\sinh{(2\pi)}} = 0.56320942\,.
\label{crlp}
\end{equation}
This value is found analytically \cite{bkwf} from a semiclassical
perturbative approach, in which the term $\gamma\,x^2y$ of the
Hamiltonian \eq{ghh} is treated as a perturbation. Using the 
perturbative trace formula given by Creagh \cite{crea} one can
extract the stabilities of the R and L orbits which in this approach
are created from the destruction of rational tori (see \cite{bkwf}
for details). To first order in the perturbation, one obtains exactly 
the correct linear approximation to $c_{RL}(\gamma)$, with the slope
\eq{crlp}, shown in the lower part of \fig{fanslope} by the 
dotted line \cite{note}.

The theoretical value of $c_{RL}(1)=6.18199717$ agrees very well with 
the value $d=6.183\pm0.001$ that was found from the numerical
stabilities of the R$_n$ and L$_{n'}$ orbits in the standard HH 
potential ($\gamma=1$) for $9\leq n,n'\leq 13$, evaluated at $e=1$.

Our result \eq{trMTRLas} obeys a known ``slope theorem''
for pitchfork bifurcations \cite{bttb,ssun,jaen1}. It states that
the slope of $\trM(e)$ of the new orbits born {\it at the bifurcation 
point} $e_n$ equals minus twice that of the parent orbit. Specifically
in the present system, it says
\be
\l{slopetheor}
\frac{\d}{\d e}\,\trMRL(e_n)
= -2\,\frac{\d}{\d e}\,\trMA(e_n)\,.
\ee
We can easily obtain the slopes of $\trMA(e)$ at $e=e_n$ from
the asymptotic result for $\trMA^{(as)}(e)$ given in \eq{trMTAf}. 
By its Taylor expansion around the asymptotic bifurcation energy
$e_n^*$ given by \eq{enas}, we find up to first order in $e-e_n^*$
\begin{equation}
\trMA^{(as)}(e) = 2 \pm c_A(\gamma)\,\frac {(e-e_n^*)}{(1-e_n^*)}
                      + {\cal O}\left[
\frac{(e-e_n^*)^2}{(1-e_n^*)^{3/2}}\right]\,.
\label{trMTAexp}
\end{equation}
The alternating sign of the linear term is
``$+$'' for the R and ``$-$'' for the L type orbit bifurcations
and thus opposite to that in \eq{trMTRLas}. 
The slope function $c_A(\gamma)$ is found to be
\bea
c_A(\gamma) = \left|\frac{\d}{\d T_A}\trMA^{(as)}(T_A,\gamma)
              \right|_{T_A=\widetilde{T}_A(e_n^*)}\!\!\!
            & = & 2\sqrt{1+2\gamma}\sqrt{|\widetilde{F}_A(\gamma)|^{2}-1}
                  \nonumber\\
            & = & \frac{2\sqrt{1+2\gamma}}{\sinh[2\pi\sqrt{1+2\gamma}]}
                  \cosh\left(\frac{\pi}{2}\sqrt{48\gamma-1}\right),
\label{slope}
\eea
where $|\widetilde{F}_A(\gamma)|$ is given in \eq{omfas} and
$\widetilde{T}_A(e)$ in \eq{Tas}. Note that $c_A(\gamma)$ does not 
depend on the bifurcation energy $e_n^*$ since $\trMA^{(as)}\,
(T_A,\gamma)$ is a periodic function of $T_A$. Comparing Eqs.\ 
\eq{crl} and \eq{slope}, we see that $c_{RL}(\gamma)=2c_A(\gamma)$
so that the theorem \eq{slopetheor} is, indeed, fulfilled with the 
correct sign.

\section{Perturbative evaluation of $\trMRL(e)$ near $e=1$}
\label{secpert}

Here we present an iterative perturbative approach for the 
calculation of the stability trace of the new orbits born 
at the bifurcation energies $\est$ of the A orbit for the GHH 
Hamiltonian \eq{ghh}, taking R orbits as example. 
This approach can be useful for Hamiltonians for which we do not 
find symmetry properties like those given in \eq{symrelmyyR} and 
\eq{symrelmyyL}, which allowed for a non-perturbative calculation
of the stability traces. 

As the small perturbation parameter we 
introduce the available energy above the bifurcation point
\be
\expeps = e-\est\,, 
\label{expeps}
\ee
which is always positive. 
The $x$ and $y$ coordinates of the new orbits, labeled $x_{po}$ and
$y_{po}$, and  the relevant elements of their monodromy matrices,
all as functions of time $t$, can be expanded in powers of 
small perturbation parameter $\expeps$:
\bea
y_{po}(t) & = & y_{\A}(t) + \expeps\,{y}_{po}^{(1)}(t) 
                 + \ldots,\qquad\;
x_{po}(t)  =  u_{po} \!\left[{x}_{po}^{(0)}(t) + 
 \expeps \,{x}_{po}^{(1)}(t) + 
        \ldots\right],
\label{xyexp}\\
\X_{ii}(t) & = &
{\X}_{ii}^{(0)}(t) + \expeps \,{\X}_{ii}^{(1)}(t) + 
 \ldots\,,\quad\;
\X_{ij}(t)  =  u_{po}\!\left[{\X}_{ij}^{(0)}(t) + 
\expeps \,{\X}_{ij}^{(1)}(t) + 
 \ldots\right] \; (i\neq j)\,,~~~
\label{Xyyxyexp}
\eea
where $i,j=x,y$.
The superscripts $^{(m)}$ indicate in an obvious manner the power 
$\expeps^m$ at which the corresponding terms appear at the $m$-th
order of the expansion. The normalization constants $u_{po}$ of 
$x_{po}(t)$ are given by 
\be\l{url}
u_R = \sqrt{\frac{e-\est}{3}}\,, \qquad 
u_L = \sqrt{\frac{e-\est}{3(1+2 \gamma y_2)}}\,.
\ee
Note that they both are proportional to $\sqrt{\expeps}$, so
that $x_{po}(t)$ goes to zero in the limit $e\to\est$. The 
solution of the equations (\ref{matrixstab0}) with the initial conditions
(\ref{initialcond}) for the stability trace $\trMRL(e)$ using
the perturbative expansions \eq{xyexp} and 
\eq{Xyyxyexp} is presented in the Appendix \ref{appc2} 
for the case of the R type orbits. The calculation for the L type
orbits is completely analogous.
The asymptotic result for $\trMR(e)$ is given in \eq{trMR2eps}.

We now compare the non-perturbative result \eq{trmmyxan} and 
the perturbative approximation \eq{trMR2eps} for the stability traces 
$\trMRL(e)$ with numerical results. Fig. \ref{fig1D} shows by
solid lines the asymptotic analytical results (\ref{trmmyxan})
for the case $\gamma=1$. They form the  ``HH fans'' with their
linear energy dependence of $\trMRL(e)$ around $e=1$, intersecting 
at the values $\trMRL(1)-2=\mp c_{RL}(1)$ with $c_{RL}(1) \approx 
6.182$ for the R and L type orbits, respectively. 
As seen from this figure, they become approximately 
symmetric with respect to the line $\trM=+2$, starting from 
$n=9$ in good agreement with the numerical results \cite{lamp}. 
Note that the linear dependence of $\trMRL(e)$ (\ref{trmmyxan}) 
is obtained up to terms of relative order $\sqrt{1-\est}$.
The perturbative result for the R type orbits (\ref{trMR2eps}) is
shown by the dashed lines, in good agreement with the analytical 
result (\ref{trmmyxan}) already for $n \geq 9$.

For further comparison with numerical results, we define the ``slope 
parameter''
\be\l{dslopepar}
d_n=|\trMRL(e=1)-2|\,,
\ee 
evaluating $\trMRL$ at the barrier ($e=1$) for a given orbit 
$\R_n$ or $\L_n$ born at the bifurcation energy $e_n$. 
As shown in \sec{sectrmrl} and Appendix \ref{app2subb}, this
parameter tends to the asymptotic limit $c_{\R\L}(\gamma)$, 
given in (\ref{crl}), for $n \to \infty$.

\tab{table1RL} shows the slope parameter $d_n$ (\ref{dslopepar})
for $7\leq n \leq 20$, evaluated for $\gamma=1$ in various
approximations; in the left part for R type orbits (odd $n$) 
and in the right part for L type orbits (even $n$). $d_n^{an}$ 
in columns 3 and 7 are the non-perturbative analytical results from 
\eq{trmmyxan}, $d_{n}^{sa}$ in column 2 represents the perturbative 
semi-analytical result \eq{trMR2eps} for the R orbits, and $d_{n}^{num}$
in columns 5 and 9 are the numerical results \cite{lamp}.
Columns 4 and 8 contain $d_n^{num*}$ obtained numerically
from solving the equations of motion (\ref{motionx}) and (\ref{motiony}) 
for the periodic orbits 
with using the FA initial conditions 
\eq{vxV2} and \eq{V1V2x2} at the top turning point \eq{toppoint},
and Eqs.\ (\ref{Xoft}) at $t=T$ for the monodromy matrix elements. 
This approximation is in good agreement with the full 
numerical results for large enough $n$, the better the larger $n$,
as seen from comparison of the 4th and 5th (and the last two) columns 
in \tab{table1RL}. The bifurcation energies for $n\geq 12$ 
were taken analytically from \tab{tabenas}. For smaller $n$, they were 
obtained by numerically solving equation $\trMA(e_n)=2$ with a 
precision better than $|\trMA(e_n)-2| \siml 10^{-9}$. 
As seen from this Table, one has  
good agreement
of the asymptotic behavior of $d_n$ of the perturbative 
$d_n^{sa}$ and even better analytical results $d_n^{an}$ 
 as compared with 
these numerical calculations. It should be noted also that the slope parameter 
(\ref{dslopepar}) of the perturbative approach \eq{trMR2eps} within the FA,
see \eq{MyyTA}, even without the correction (\ref{y2eps}) 
to the periodic orbit $y_{\A}(t)$, 
is in rather good agreement with the numerical results presented in 
\tab{table1RL}, especially for asymptotically large $n$, with a precision 
better than 5\%. However, the second correction in \eq{trMR2eps} above 
the FA improves essentially the slope parameter (\ref{dslopepar}) 
in this asymptotic region.
As noted above, the asymptotic values of the perturbative 
$d_{n}^{sa}$ and the non-perturbative $d_n^{an}$, 
as well as the numerical FA result for $d_n^{num*}$,  
all converge sufficiently rapidly to the asymptotic analytical 
number $ c_{RL}(1)=6.18199717$ given by Eq.\ \eq{crl}, in 
line with the analytical convergence found above from \eq{trmmyxanas}.

Fig. \ref{figghhan} shows good agreement between the analytical 
\eq{trmmyxan}, semi-analytical \eq{trMR2eps}
and numerical solving the GHH equations \eq{motionx},
and \eq{motiony} for classical periodic orbits
and \eq{Xoft} for the monodromy matrix with FA initial conditions for
 $L_{12}$ and $R_{13}$ as examples. Both these curves agree very well
with the asymptotic analytical slopes $c_{RL}(\gamma)$ 
within a rather wide interval of $\gamma$ even for not too large $n$
of the orbits
mentioned above. This comparison is improved with increasing $n$,
the better the larger $n$, which gives a numerical confirmation of
the analytical convergence of the $\trMRL(e,\gamma)$ \eq{trmmyxan}   
to the asymptotic $c_{RL}(\gamma)$ \eq{crl} at 
the barrier $e = 1$ for any $\gamma$. For larger $\gamma$, 
one needs larger $n$ in order to obtain convergence of all 
the compared curves.

\section{Summary and conclusions}

In this paper we have investigated the bifurcation cascades
of the linear A orbit in a class of generalized H\'enon-Heiles
(GHH) potentials. We were able to derive analytical expressions for
the stability traces $\trMA(e)$ of the A orbit and $\trMRL(e)$
of the R and L orbits bifurcating from it as functions of the
energy, which are asymptotically valid for energies close to 
the saddle at $e=1$, i.e., in the limit where the bifurcations 
energies $e_n$ approach the saddle: $e_n\to 1$. Our results
confirm analytically the empirical numerical properties of the
``H\'enon-Heiles fans'' that are formed by the asymptotically
linear intersection of the functions $\trMRL(e)$ at $e=1$, as given in 
Eq.\ \eq{trMTRLas}. We found good agreement of our alternative 
non-perturbative and perturbative asymptotic results for
$\trMRL(e)$ with the numerical results. As a bonus, we have also 
obtained asymptotically exact expressions for the bifurcation energies
$e_n$ of the A orbit in the GHH system, given in Eq.\ \eq{enas}.
Our results can be interpreted in the sense that the 
non-integrable, chaotic GHH Hamiltonian becomes approximately 
integrable locally at the barrier, i.e., for $e=1$.

Both our approaches may be useful, also for more general Hamiltonians,
for semiclassical calculations of the Gutzwiller trace formula for the 
level density \cite{gutz}, extended to bifurcation cascades with the 
help of suitable normal forms and corresponding uniform approximations 
\cite{ozob,ssun}. A normal form with uniform approximation for two
successive pitchfork bifurcations has been derived and successfully
applied to the HH system in \cite{jkmb}. In future research, we hope
to generalize the normal form theory to infinitely dense bifurcation 
sequences with the help of the results of \cite{maf,jkmb} and the
theory of Fedoryuk \cite{fedorpr,fedorbook}. Hereby the
``HH fan'' phenomenon for the stability traces might be useful. 

\bs
\bs

\begin{acknowledgments}

S.N.F.\ and A.G.M.\ acknowledge the hospitality at Regensburg 
University during several visits and financial support by the 
Deutsche Forschungsgemeinschaft (DFG) through the graduate 
college 638 ``Nonlinearity and Nonequilibrium in Condensed 
Matter''.

\end{acknowledgments}

\appendix

\section{Asymptotic evaluation of $\trMA(e)$ for $e\to 1$}
\label{app1}

To obtain the stability matrix $\MA$ for the orbit A, we have to solve
the linearized equation of motion \eq{hill}\ for small perturbations 
around the orbit in the perpendicular direction. Since the A orbit
moves along the $y$ axis, we have $q=x$, $p=\dot{x}$ and \eq{hill} becomes
\eq{motionx} which is already linear in $x$. We thus find $\MA$ from the 
non-periodic solutions of \eq{motionx} with small initial values $x_0=x(t=0)$, 
${\dot x}_0={\dot x}(t=0)$. Let us denote these solutions by 
$x(t;x_0,{\dot x}_0)$. The elements of $\MA$ (we omit the subscript 
``$\perp$A'' for simplicity) are then given by
\begin{equation}
\M_{qq}=\lim_{x_0\to0}{\frac{x(T_A;x_0,0)}{x_0}}\,, \qquad
\M_{qp}= \lim_{\dot{x}_0\to0}{\frac{x(T_A;0,\dot{x}_0)}{ \dot{x}_0}}\,,
\label{M11}
\end{equation}
\begin{equation}
\M_{pq}=\lim_{x_0\to0}{\frac{\dot{x}(T_A;x_0,0)}{x_0}}\,, \qquad
\M_{pp}=\lim_{\dot{x}_0\to0}{\frac{\dot{x}(T_A;0,\dot{x}_0)}{ \dot{x}_0}}\,.
\label{M22}
\end{equation}

We could not find exact analytical solutions of \eq{motionx} using 
the exact function $y_A(t)$ \eq{yAt} for the A orbit, for which
\eq{motionx} becomes the Lam\'e equation. Only at the bifurcation 
energies $e_n$, one of its solutions is a periodic Lam\'e function 
which has known expansions \cite{erde}. For the non-periodic solutions,
no expansions could be found in the literature. We can, however, solve 
\eq{motionx} if we instead of the exact $y_A(t)$ use the approximation 
${\tilde y}_A(t)$ given in 
\eq{yTheta}, which becomes exact in the asymptotic limit $e\to1$, 
and for which \eq{motionx} can be reduced to the Legendre equation as 
shown below. We proceed separately for the two time intervals $0\leq t 
\leq t_2$ and $t_2\leq t \leq T_A$, as specified after \eq{fact}:

\ms

\noindent
a) $0\leq t \leq t_2$: Solve the equation
\begin{equation}
 \ddot{x}_1(t)+[1+2\gamma Y_1(t)]\, x_1(t)=0\,,
\label{x1eq}
\end{equation}
with the initial conditions
\begin{equation}
 x_1(0)= x_0=0\,, \qquad \dot{x}_1(0) =\dot{x}_0 \to0\,,
\label{1initial}
\end{equation}
and obtain $x_1(t_2)$.

\ms

\noindent
b) $t_2\leq t \leq T_A$: Solve the equation
\begin{equation}
\ddot{x}_2(t)+[1+2\gamma Y_2(t)]\, x_2(t)=0\,,
\label{x2eq}
\end{equation}
with the initial conditions
\begin{equation}
  x_2(t_2)=x_1(t_2)\,, \qquad \dot{x}_2(t_2)=\dot {x}_1(t_2)\,,
\label{2initial}
\end{equation}
and obtain $x_2(T_A)$.

To do so, we transform equations \eq{x1eq}, \eq{x2eq} by 
defining the following variables:
\be
z_1=z\,, \qquad z_2=z-2\,\K(\kappa)\,. 
\ee
Then, the equations 
(\ref{x1eq}), (\ref{x2eq}) can be written 
compactly as:
\begin{equation}
 \frac{{\rm d}^2}{{\rm d} z^2} x_i(z_i)+[B+A\tanh^2(z_i)]\, x_i(z_i)=0\,, 
\qquad  (i=1,2)
\label{ten}
\end{equation}
where
\begin{equation}
B= \frac{6(1+2\gamma y_1)}{(y_3-y_1)}\,, \qquad
A=\frac{12\gamma (y_2-y_1)}{(y_3-y_1)}\,.
 \label{eleven}
\end{equation}
We next go over to the new variables 
\be
s_i=\tanh(z_i)\,. \qquad   (i=1,2)
\ee
Then \eq{ten} is transformed into the Legendre equation:
\begin{equation}
(1-s_i^2)\,\frac{{\rm d}^2}{{\rm d} s^2} x_i(s_i)
-2s_i\,\frac{{\rm d}}{{\rm d} s} x_i(s_i)+\left[\nu(\nu +1)-\frac{\mu^2}{1-s^2_i}\right]
\!x_i(s_i)=0\,, \qquad (i=1,2)
\label{legendre}
\end{equation}
with
\begin{equation}
\mu =i\sqrt{A+B}\,, \qquad    \nu= (-1+i\sqrt{4A-1})/2\,.
 \label{munu}
\end{equation}
The Legendre equation 
(\ref{legendre}) has the solution
\begin{equation}
 {x}_i(s_i)= C_{1i}\Pl_\nu^\mu(s_i) + C_{2i}\Ql_\nu^\mu(s_i)\,, \qquad  (i=1,2),
\label{PQ}
\end{equation}
where $\Pl_\nu^\mu(s)$ and $\Ql_\nu^\mu(s)$ are the associated Legendre functions 
of first and second kind, respectively, with real argument $-1\leq s\leq +1$ 
(see \cite{grry}). 
The initial conditions 
(\ref{2initial}) for $x_2(s_2)$ have the form
\begin{equation}
x_2(-s_K)=x_1(s_K)\,, \qquad \quad
\Bigl(\frac{\d x_2(s_2)}{\d s_2}\Bigr)_{s_2=-s_K}=
\Bigl(\frac{\d x_1(s_1)}{\d s_1}\Bigr)_{s_1=s_K}\,,
 \label{3initial}
\end{equation}
where
\begin{equation} 
s_K=\tanh\K(\kappa) = \tanh(a_\kappa T_A/2)\,.
\label{sk}
\end{equation}

Solution (\ref{PQ}) 
of equation \eq{legendre} for $i=1$ with the initial conditions 
(\ref{1initial}) 
yields the following 
expressions for the coefficients $C_{11}$ and $C_{21}$:
\begin{equation}
 C_{11}= \dot{x}_0\, D^{\mu}_{\nu}(s_F)\, \Ql^{\mu}_{\nu}(s_F)\,,\qquad
 C_{21}= -\dot{x}_0\, D^{\mu}_{\nu}(s_F)\, \Pl^{\mu}_{\nu}(s_F)\,,
\end{equation}
with
\begin{equation}
s_F=\tanh F(\varphi,\kappa)\,, \qquad
D^{\mu}_{\nu}(s_F)= \left[a_\kappa(1-s_F^2)\, W_\nu^\mu (s_F)\right]^{-1}.
 \label{D1}
\end{equation}
Here $W_\nu^\mu(s_F)$ is the Wronskian
\begin{equation}
 W_\nu^\mu(s) \equiv 
W\{\Ql^{\mu}_{\nu}(s),\Pl^{\mu}_{\nu}(s)\}= \Ql^{\mu}_{\nu}(s)
\frac{{\rm d}}{{\rm d} s}{\Pl^{\mu}_{\nu}}(s)
  -\Pl^{\mu}_{\nu}(s) \frac{{\rm d}}{{\rm d} s}{\Ql^{\mu}_{\nu}}(s)\,.
\label{Wro}
\end{equation}
Analogously, we solve equation (\ref{legendre}) for $i=2$ with the
initial conditions 
(\ref{3initial}) and obtain for $C_{12}$ and $C_{22}$ 
the following expressions: 
\begin{equation}
 C_{12} = \dot{x}_0\, D^{\mu}_{\nu}(s_F) \, \frac{a_1\Ql^{\mu}_{\nu}(s_F)
                             +a_2\Pl^{\mu}_{\nu}(s_F)}{W_\nu^\mu(s_K)}\,,
\label{C12}
\end{equation}
\begin{equation}
 C_{22} =\dot{x}_0\, D^{\mu}_{\nu}(s_F) \, \frac{a_3\Ql^{\mu}_{\nu}(s_F)
                             +a_4\Pl^{\mu}_{\nu}(s_F)}{W_\nu^\mu(s_K)}\,.
\label{C22}
\end{equation}
The coefficients $a_i$ here are:
$$a_1= \Ql^{\mu}_{\nu}(-s_K)\frac{{\rm d}}{{\rm d} s}{\Pl^{\mu}_{\nu}}(s_K)-
 \Pl^{\mu}_{\nu}(s_K)\frac{{\rm d}}{{\rm d} s} {\Ql^{\mu}_{\nu}}(-s_K)\,,$$
$$a_2= \Ql^{\mu}_{\nu}(s_K)\frac{{\rm d}}{{\rm d} s}{\Ql^{\mu}_{\nu}}(-s_K)-
 \Ql^{\mu}_{\nu}(-s_K)\frac{{\rm d}}{{\rm d} s} {\Ql^{\mu}_{\nu}}(s_K)\,,$$
$$a_3= \Pl^{\mu}_{\nu}(s_K)\frac{{\rm d}}{{\rm d} s}{\Pl^{\mu}_{\nu}}(-s_K)-
 \Pl^{\mu}_{\nu}(-s_K)\frac{{\rm d}}{{\rm d} s} {\Pl^{\mu}_{\nu}}(s_K)\,,$$
\be
a_4= \Pl^{\mu}_{\nu}(-s_K)\frac{{\rm d}}{{\rm d} s}{\Ql^{\mu}_{\nu}}(s_K)-
 \Ql^{\mu}_{\nu}(s_K)\frac{{\rm d}}{{\rm d} s} {\Pl^{\mu}_{\nu}}(-s_K)\,.
\label{ai}
\ee
Using $x_2$ in (\ref{PQ}) at $t=T_A$ and (\ref{C12}), 
(\ref{C22}) for the coefficients $C_{12}$, $C_{22}$, we obtain the 
following expression for $\M_{pp}$ defined in (\ref{M22}):
\bea
\M_{pp}&=&\left[W_\nu^\mu(s_F)W_\nu^\mu(s_K)\right]^{-1}\left[
a_1 \Ql^{\mu}_{\nu}(s_F)\frac{{\rm d}}{{\rm d} s}{\Pl^{\mu}_{\nu}}(s_F) + a_2
\Pl^{\mu}_{\nu}(s_F)\frac{{\rm d}}{{\rm d} s}{\Pl^{\mu}_{\nu}}(s_F)\right. 
\nonumber\\
&+& \left. a_3
\Ql^{\mu}_{\nu}(s_F)\frac{{\rm d}}{{\rm d} s}{\Ql^{\mu}_{\nu}}(s_F) + a_4
\Pl^{\mu}_{\nu}(s_F)\frac{{\rm d}}{{\rm d} s}{\Ql^{\mu}_{\nu}}(s_F)\right].
\label{M22F}
\eea
To calculate $\M_{qq}$ defined in (\ref{M11}), we solve equations (\ref{x1eq}),
(\ref{x2eq}) with the initial conditions
\begin{equation}
x_1(0)=x_0\to0\,, \qquad \dot{x}_1(0)=\dot{x}_0=0\,,
\label{4initial}
\end{equation}
and then take into account the condition (\ref{2initial}). Using the
same steps as for $\M_{pp}$, we obtain $\M_{qq}$ in the following form:
\bea
\M_{qq}&=&- \left[W_\nu^\mu(s_F) W_\nu^\mu(s_K)\right]^{-1}\left[
a_1 \Pl^{\mu}_{\nu}(s_F)\frac{{\rm d}}{{\rm d} s}{\Ql^{\mu}_{\nu}}(s_F) + a_2
\Pl^{\mu}_{\nu}(s_F)\frac{{\rm d}}{{\rm d} s}{\Pl^{\mu}_{\nu}}(s_F) \right.
\nonumber\\
&+& \left.a_3
\Ql^{\mu}_{\nu}(s_F)\frac{{\rm d}}{{\rm d} s}{\Ql^{\mu}_{\nu}}(s_F) + a_4
\Ql^{\mu}_{\nu}(s_F)\frac{{\rm d}}{{\rm d} s}{\Pl^{\mu}_{\nu}}(s_F)\right]. 
\label{M11F}
\eea
Using the following explicit expression for the Wronskian (\ref{Wro}),
\begin{equation}
 W^{\mu}_{\nu}(s)=
 \frac{1}{(s^{2}-1)}\frac{\Gamma(1+\nu+\mu)}{\Gamma(1+\nu-\mu)}\,,
\label{Wro2}
\end{equation}
we now find for the sum of $\M_{qq}$ and $\M_{pp}$ 
\begin{equation}
\trMA(e)=2\,(s_K^{2}-1)\,\frac{\Gamma(1+\nu-\mu)}{\Gamma(1+\nu+\mu)}
           \left[\Ql^{\mu}_{\nu}(-s_K)\frac{{\rm d}}{{\rm d} s}{\Pl^{\mu}_{\nu}}(s_K)
           -\Pl^{\mu}_{\nu}(-s_K)\frac{{\rm d}}{{\rm d} s}{\Ql^{\mu}_{\nu}}(s_K)\right]\!.
\label{trMTA2}
\end{equation}
Note that the energy dependence comes through the quantities $\mu$, $\nu$
given in \eq{munu} and $s_K$ in \eq{sk} via the turning points $y_i(e)$ 
given in \eq{yy}. As must be expected, the result (\ref{trMTA2}) does not 
depend on the initial point $y_0$. 

We recall that the result 
(\ref{trMTA2}) has been obtained using the
approximation 
(\ref{yTheta}) for the function $y_A(t)$, which is based
on the asymptotic expression (\ref{fact}) for the Jacobi elliptic
function \sn$(z,\kappa)$, valid in the limit $\kappa\to1$.
We can therefore simplify the above result by taking asymptotic 
limits, valid for $e\to1$, of the quantities appearing in \eq{trMTA2}.
Since we have omitted the next-to-leading correction to \eq{fact}, 
it is consistent to keep only the leading asymptotic terms. (An
evaluation of all next-to-leading order corrections would lead
outside the scope of this paper.)

Using the asymptotic forms of the Legendre functions through 
hypergeometric series (cf.\ \cite{grry}, Eqs.\ 8.704, 8.705, 
and 8.737), we obtain for the leading term in (\ref{trMTA2}) 
the intermediate result
\begin{equation}
\trMA(e) \approx 2\,\Re \left(e^{-a_\kappa T_A\mu}F_A\right),
\label{trMTA3}
\end{equation}
where the function $F_A(\gamma)$ is defined by
\begin{equation}
F_A(\gamma)=\frac{\mu\pi}{\sin^2(\mu\pi)}\,
            \frac{\Gamma(1+\nu+\mu)\,\sin[(\nu+\mu)\pi]}
            {\Gamma(1+\nu-\mu)\,\Gamma^2(1+\mu)}
           =|F_A(\gamma)|\,e^{i\Phi_A(\gamma)}\,.
\label{FA}
\end{equation}
Here the period $T_A$ and the quantities $a_\kappa$ in \eq{qmod},
and $\mu,\nu$ given in \eq{munu} still depend on the energy $e$.
Now, for $e\to1$, all quantities in \eq{trMTA3} except $T_A(e)$
have finite limits, easily found from the limiting turning points
$y_1\to-1/2$, $y_2\to1$, $y_3\to1$. In particular, we get the limits:
\begin{equation}
a_\kappa\to 1/2\,, \quad
\mu \to 2i\sqrt{1+2\gamma}\,, \quad   
\nu \to \frac12\left(-1+i\sqrt{48\gamma-1}\right). \qquad (e\to 1)
 \label{munu2}
\end{equation}
The limit of $F_A(\gamma)$ will be denoted by $\widetilde{F}_A(\gamma)$
and its limiting phase by $\widetilde{\Phi}_A(\gamma)$
\be
F_A(\gamma) \; \to \; \widetilde{F}_A(\gamma)
=  |\widetilde{F}_A(\gamma)|\,e^{i\widetilde{\Phi}_A(\gamma)}\,. \qquad (e\to1)
\label{fi}
\ee
Its modulus can be given analytically as
\be
|\widetilde{F}_A(\gamma)| =
\frac{\sqrt{\cosh(4\pi\sqrt{1+2\gamma})+\cosh(\pi\sqrt{48\gamma-1})}}
     {\sqrt{2}\,\sinh[2\pi\sqrt{1+2\gamma}]}. 
\label{omfas}
\ee
We discuss only positive values of $\gamma$ here; for $\gamma < 1/48$, 
the function $\cosh{(\pi\sqrt{48\gamma-1})}$ becomes equal to 
$\cos{(\pi\sqrt{1-48\gamma})}$. The phase $\widetilde{\Phi}_A(\gamma)$
is defined through Eqs.\ \eq{FA} and \eq{fi}; 
it turns out to be negative for all $\gamma>0$.

Using the above limits, we finally get from \eq{trMTA3} the asymptotic 
expression for $\trMA(e)$ given in Eqs.\ \eq{trMAeas} and \eq{trMTAf}
of \sec{secstab}.

\section{Asymptotic evaluation of $\trMRL(e)$ for $e\to 1$}
\l{app2}

As mentioned in \sec{sechill}, the stability matrix $\Mt$ of a
periodic orbit in a
two-dimensional system is found by linearization of the equations of 
motion in the phase-space variables $\bgrk{\xi}_\perp = (q,p)$
transverse to the orbit. For the R and L orbits, which have
curved shapes that are only known numerically above their
bifurcation energies, we have no way of determining the
variables $(q,p)$ analytically. We are therefore forced to
evaluate the diagonal elements of the full monodromy matrix M, 
in order to find $\trM=\;$tr\,M$-2$ for these orbits.
For their calculation, we exploit some symmetry relations which are 
valid when the starting point at $t=0$ of a periodic orbit is chosen 
to be the upper turning point in the direction of the A orbit, i.e., 
along the $y$ axis:
\be
y(t=0)=y_{max}\,,\qquad x(0)=0\,.
\label{toppoint}
\ee
We first present these relations for the A orbit and then for the R 
and L orbits.

\subsection{Symmetry relations for elements of monodromy matrix M}
\l{app2suba}

\subsubsection{Diagonal elements for A orbit}
\l{app2suba1}

For the straight-line librating orbit A, we have $\xip=y$,
$\dxip=\doty$, and hence we may apply immediately \eq{parid}. 
For the calculation of the elements $\M_{xx}$ and $\M_{\dotx\dotx}$,
we note that the differential equations for $\X_{xx}(t)$ and
$\X_{\dotx\dotx}(t)$ contained in \eq{Xoft} decouple for the
A orbit. Writing them at the time $t=T$, where $T$ is the period
of the A orbit, they can be combined into the following
second-order differential equations for $\M_{xx}(T)$ and
$\M_{\dotx \dotx}(T)$ as functions of the variable $T$:
\begin{equation}
\frac{\d^2}{\d T^2} \M_{xx}(T) + V_{xx}(T)\,\M_{xx}(T)=0\,,
\label{MxxeqTA}
\end{equation}
\begin{equation}
\frac{\d^2}{\d T^2} \M_{\dotx \dotx}(T) + V_{xx}(T)\,\M_{\dotx \dotx}(T) =
V_{xxy}(T) \,\M_{x \dotx}(T)\,\doty_A(T)\,,
\label{MdxdxeqTA}
\end{equation}
where the subscripts of $V$ denote its corresponding (successive) partial 
derivatives. With the special choice of the starting point \eq{toppoint},
which for the A orbit becomes $y_A(0)=y_2$, see \eq{yAt},
we have $\doty_A(0)=\doty_A(T)=0$ and the two equations for $\M_{xx}(T)$ 
and $\M_{\dotx\dotx}(T)$ become identical. For solving them uniquely, 
two boundary conditions are sufficient. Since both quantities become
unity at bifurcations, may we choose two successive periods
$T=T_n=T(e_n)$ and $T=2T_n$ at the bifurcation energy $e=e_n$
to impose the boundary condition
\begin{equation} 
\M_{xx}(T_n)=\M_{\dotx \dotx}(T_n)=1,\qquad\qquad
\M_{xx}(2T_{n})=\M_{\dotx \dotx}(2T_{n})=1\,.
\label{mxxdxdxbifT}
\end{equation}
This ensures the uniqueness of the solutions, so that we obtain
the result
\be
\M^{(A)}_{xx}=\M^{(A)}_{\dotx \dotx}\,,
\l{symrelxA}
\ee
which holds at arbitrary periods $T$ and hence at arbitrary energies $e$.

\subsubsection{Diagonal elements for R and L orbits}
\l{app2suba2}

For the R orbits born at the successive bifurcation energies
$e_n$, we have $\xip=x$, $\dxip=\dotx$ at the starting point
\eq{toppoint}, while $y$ is the coordinate perpendicular to the 
orbit and one may apply \eq{parid}. 
To obtain the elements $\M_{yy}$ and $\M_{\doty \doty}$ of the R
orbits at the starting point (\ref{toppoint}), we may use the ``frozen
approximation'' (FA) for the $y$ motion of these orbits (cf.\ 
\cite{mbgu,lamp}) which is taken to be that of the A orbit, $y_R(t)\approx 
y_A(t)$, so that the starting point is at $y_{max}=y_2$. This corresponds
strictly to the lowest order of the perturbation expansion in the
small parameter $\epsilon=e-e_n$. 
Then, the velocity $v_x$ of their $x$ motion close to $e=e_n$ is 
proportional to $\sqrt{e-e_{n}}$ as given in \eq{vxV2} below.
For the functions $\M_{yy}(T)$ and $\M_{\doty \doty}(T)$, equations 
analogous to \eq{MxxeqTA} and \eq{MdxdxeqTA} hold, but with the
subscripts $x,\dotx$ and $y,\doty$ exchanged and $\doty_A$ replaced
by $\dotx_R$, $T$ now being the period of an R orbit; boundary
conditions analogous to \eq{mxxdxdxbifT} apply. Hence we can conclude 
that in the limit $e\to 1$, where $\epsilon=e-e_n$ becomes small,
the following approximate symmetry relation holds for the R orbits:
\be
\M^{(R)}_{yy}\approx\M^{(R)}_{\doty \doty}\,. \qquad (e\to 1) 
\l{symrelyR}
\ee

For the L orbits, the situation is slightly more difficult: their
upper turning point does not lie on the $y$ axis, nor do they
reach or leave their turning point in the $x$ direction.
However, the $x$ coordinate at the turning point is proportional
to $\sqrt{e-e_n}$ close to their bifurcation energy $e_n$.
Furthermore, the coordinate system $(x,y)$ can be rotated such 
that the L orbits move in the rotated $x$ direction at their
upper turning points, and the diagonal elements of $M$ are not
changed under this rotation. Thus, the relation 
\eq{symrelyR} is, to leading order in $\epsilon=e-e_n$, also
found to hold for the L orbits: 
\be
\M^{(L)}_{yy}\approx\M^{(L)}_{\doty \doty}\,. \qquad (e\to 1) 
\l{symrelyL}
\ee

\subsubsection{Relations of diagonal to non-diagonal elements}
\l{app2suba3}

Other symmetry relations can be obtained by taking the variational 
(partial) derivatives of the energy conservation equation at $t=T$: 
\be
H\left[x(T),y(T),\dotx (T),\doty (T)\right]=E\,,
\l{enconstyt}      
\ee
with respect to the initial variables, e.g., $y(0)$ and $\doty(0)$. 
Differentiating (\ref{enconstyt}) in $y(0)$ and $\doty(0)$ and
applying the definition of the monodromy matrix elements 
(\ref{defX}), one has
\bea\l{eqsymrel}
V_x\;\M_{xy} &+& V_y\;\M_{yy} 
+ \dotx \M_{\dotx y}+\doty \M_{\doty y}=V_y,\nonumber\\
V_x\;\M_{x\doty} &+&
V_y\;\M_{y\doty} 
+ \dotx \M_{\dotx \doty}+\doty \M_{\doty \doty}=\doty,
\eea
where 
\be\l{vxvy}
V_x=\frac{\partial V}{\partial x}=
x(1+2 \gamma y), \qquad
 V_y=\frac{\partial V}{\partial y}=
y(1-y)+ \gamma x^2,
\ee
according to the GHH Hamiltonian (\ref{ghh}). All coefficients 
in front of the monodromy matrix elements are taken at the
periodic orbit under consideration:  $x \equiv x_{po}(T)=x_{po}(0)$, 
$y \equiv y_{po}(T)=y_{po}(0)$, etc. From \eq{eqsymrel} at the 
starting point \eq{toppoint} for the R orbit, which in the FA is
$y_R(0)=y_2$, $x_R(0)=0$,  one finds with $\doty_R(0)=0$
\be\l{myydxdyR}
\M_{yy} = 1 -\frac{v_x}{V_2}\, \M_{\dotx y}, \qquad\qquad
\M_{y \doty}=-\frac{v_x}{V_2}\, \M_{\dotx \doty},
\ee
where 
\be\l{vxV2}
v_x=\dotx(0) \approx \sqrt{\frac{e-e_n}{3}}, \qquad\qquad
V_2=y_2(1-y_2) \approx \sqrt{\frac{1-e_n}{3}},
\ee  
see \eq{vxvy}. 
The results in \eq{vxV2}, as well as all
approximate relations given below, are valid in the FA in the 
limit $e,e_n \to 1$ (with $e>e_n$) and are correct to leading 
order in $\sqrt{1-e_n}$. From this one obtains the two approximate 
symmetry relations
\be\l{mdxydxdymyxydy}
\M^{(R)}_{\dotx y}\approx\M^{(R)}_{\dotx \doty},\qquad\qquad
\M^{(R)}_{y \doty}\approx\frac{v_x}{V_2}\,\M^{(R)}_{y x}.
\ee
The first relation follows from the identical differential equations 
for the functions  $\M_{\dotx y}(T)$ and $\M_{\dotx \doty}(T)$
at the turning point \eq{toppoint} of the R orbits:
\bea\l{eqmdxymdxdy}
\ddot{\M}_{\dotx y}(T)+[1+2 \gamma y_{\R}(T)]\,M_{\dotx y}(T)
= -2 \gamma\; \dotx_{\R}(T)\,\M_{yy}(T)\,,
\nonumber\\
\ddot{\M}_{\dotx \doty}(T)+[1+2 \gamma y_{\R}(T)]\, M_{\dotx \doty}(T)
= -2 \gamma \;\dotx_{\R}(T)\,\M_{\doty \doty}(T)\,,
\eea
according to \eq{symrelyR}, and their zero initial values at $e=e_n$.
The second symmetry relation in \eq{mdxydxdymyxydy} can be proved
directly through their definitions
(\ref{defX}),
\be\l{defmydyyx}
\frac{\M_{y \doty}}{\M_{y x}}= \frac{\delta x(0,e)}{\delta \doty(0,e)}=
\left(\frac{\delta x(0,e)/\delta e}{\delta 
 \doty(0,e)/\delta e}\right)_{e \rightarrow e_n}, 
\ee
where we write explicitly the energy dependence of the trajectory 
 $\{x(t,e)$, $y(t,e)\}_{po}$
owing to the initial conditions besides of the time dependence  considered
above.
By employing the condition at the R top point, we find
\be\l{dycond}
\doty_{\R}(T(e),e) \equiv 0 = \doty_{\R}(T(e_n),e_n) + (e-e_n) 
\!\left[\ddot{y}_{\R} T'(e_n) + 
\frac{\partial \doty_{\R}(T,e_n)}{\partial e}\right].
\ee
We used here the $y$ equation of motion (\ref{motiony}) in order to 
obtain the derivative in the denominator of the r.h.s. in \eq{defmydyyx}. 
For the derivative in the numerator we may use the FA near the saddle
energy, $\delta x(0,e_n)/\delta e = \dotx(0) T'(0)$,
because the main energy dependence is coming through 
the period $T(e)$ in the argument of $x_{po}(T,e)$.
Finally, from \eq{myydxdyR} and \eq{mdxydxdymyxydy} one arrives at 
two other useful approximate symmetry relations
\be\l{symrelmyyR}
\M^{(R)}_{yy}\approx 1+ \M^{(R)}_{y \doty}\approx 1 + 
\frac{v_x}{V_2}\,\M^{(R)}_{yx}\,.
\ee
In an analogous way, from \eq{eqsymrel}, one directly derives the following 
symmetry relations for the L orbits accounting for their different initial 
conditions at the top (turning) point,
 $\doty_\L(0)=\dotx_\L(0)=0$, $y_\L(0)=y_2$, 
$x_\L(0)=x_2$ (cf.\ \cite{lamp}),
\be\l{symrelmyyL}
\M^{(L)}_{yy} \approx 1 + \M^{(L)}_{\doty y} \approx 1 - 
\frac{V_1}{V_2}\,\M^{(L)}_{xy}\,,
\ee
where 
\be\l{V1V2x2} 
 V_1=x_2 (1+2 \gamma y_2), \qquad x_2 \approx 
\sqrt{\frac{e-e_n}{3(1+ 2 \gamma y_2)}}\,,\qquad
V_2=y_2(1-y_2)+\gamma x_2^2 \approx \sqrt{\frac{1-e_n}{3}}\,,
\ee
where the FA has been used.

Other symmetry relations between monodromy matrix elements can be obtained 
in a similar way. In particular, one obtains the following structure
of the stability matrix for both R and L orbits, 
\begin{equation}
{\rm M}_{\perp R,L} = \left({\M_{yy} \qquad \M_{y \doty}\atop{\M_{\doty y} 
\qquad \M_{\doty \doty}}}\right)_{R,L} \approx 
\left({\M_{yy} \qquad \M_{y y} \mp 1\atop{\M_{y y} \pm 1
\qquad \M_{yy}}}\right)_{R,L}\,,
\label{Mstab}
\end{equation}
where the upper sign holds for R and the lower for L orbits.
 
All approximate symmetry relations (\ref{symrelyR}), (\ref{myydxdyR}), 
(\ref{mdxydxdymyxydy}), (\ref{symrelmyyR}) and (\ref{symrelmyyL}) 
and the structure (\ref{Mstab}) of the stability matrix have been 
checked by explicit numerical calculations, solving \eq{Xoft}
in the FA with the given starting conditions. They become the more 
accurate the closer the bifurcation energies $e_n$ are to the saddle 
energy $e=1$.

In conclusion, we need not calculate those three quarters of the 
matrix $\X(t)$ in which the indices $\dot x$ or $\dot y$ appear.
The coupled differential equations for the remaining elements of 
$\X(t)$ are 
\bea
\ddot{\X}_{xx}(t) + [1+2 \gamma y_{po}(t)]\,\X_{xx}(t) & = &
-2\gamma\; x_{po}(t)\,\X_{yx}(t)\,,\nonumber\\
\ddot{\X}_{yx}(t) + [1-2 y_{po}(t)]\,\X_{yx}(t)  & = & 
-2\gamma\;x_{po}(t)\,\X_{xx}(t)\,,\nonumber\\
\ddot{\X}_{yy}(t) + [1-2 y_{po}(t)]\,\X_{yy}(t) & = &
-2\gamma\;x_{po}(t)\,\X_{xy}(t)\,,\nonumber\\
\ddot{\X}_{xy}(t) + [1+2 \gamma y_{po}(t)]\,\X_{xy}(t)  & = & 
-2\gamma\;x_{po}(t)\,\X_{yy}(t)\,,\l{matrixstab0}
\eea
to be solved with the initial conditions
\bea\l{initialcond}
\X_{xx}(0) =\X_{yy}(0)=1\,,\qquad \dot{\X}_{xx}(0)=\dot{\X}_{yy}(0)=0\,\,,
\nonumber\\
\X_{xy}(0)=\X_{yx}(0)=0\,,\qquad \dot{\X}_{xy}(0)=\dot{\X}_{yx}(0)=0\,.
\eea
In the equations (\ref{matrixstab0}), the functions
$y_{po}(t)$ and $x_{po}(t)$ describe the $y$ and $x$ 
motion of the periodic R and L orbits, respectively, born at the 
bifurcations.

By using the relations (\ref{parid}) for $r_\parallel=x,$ 
$\dot{r}_\parallel=\dot{x}$ 
and (\ref{symrelyR}), (\ref{symrelyL}),
one has the stability matrix trace 
of $\Mt$ for the R and L orbits,
\be\l{tracR1}
\trM \approx 2\,\left[\X_{xx}(T_{\A})+\X_{yy}(T_{\A})\right] -2\,
=2\,\left(\M_{xx}+\M_{yy}\right) -2 = 2 \M_{yy}\, ,
\ee
where $T_{\A}$ is the period of A orbit, $T_{\A}=T_{\A}(e)$, 
taken in the FA 
at the bifurcation energy, $e=\est$, $\M_{ij}=\X_{ij}(T_{\A})$
($\M_{xx} = 1$). 

\subsection{Analytical asymptotic expressions for $\trMRL(e)$}
\l{app2subb}

To solve the system of equations (\ref{matrixstab0}), we have to
specify the functions $x_{po}(t)$. In the asymptotic limit $e \to 1$, we
can use the FA in which $y_{po}(t) \approx y_{\A}(t)$.
The stability equation for the R and L orbits then is
\begin{equation}
\ddot{x}_{po}(t)+[1+2 \gamma y_A(t)]\,x_{po}(t)=0\,, \qquad (po=R,L)
\label{XAeq}
\end{equation}
with the initial conditions $x_R(0)=0$, $\dot{x}_R(0)=v_x$ and $x_L(0)=x_2$,
$\dot{x}_L(0)=0$.
As discussed in \cite{lamp}, \eq{XAeq} with the exact $y_A(t)$ given in 
\eq{yAt} is the Lam\'e equation, whose periodic solutions are the
periodic Lam\'e functions. However, for $e\to1$ we may replace the 
sn function in \eq{yAt} by its asymptotic form given in \eq{fact}:
\begin{equation}
y_{\A}(t) \approx y_1+(y_2-y_1)s^2(t),\qquad\qquad
s(t) = \tanh\left[a_{\kappa}t-\K(\kappa)\right],
\label{y_Ath}
\end{equation}
and transform the equation \eq{XAeq} to the Legendre equation 
(\ref{legendre}), replacing $s_i \to s$ and $x_i(s_i) \to x_{po}(s)$,
with $s(t)$ given in \eq{y_Ath}. We then obtain the
$x_{po}(t)$ in terms of the Legendre functions as
\be\l{x_Ath}
x_\R(t)= v_x\,\Phi_{+}\left(s_K,s\right)\!/(a_{\kappa}\overline{W}_{+})\,,
         \qquad\qquad
x_\L(t)= x_2\,\Psi_{+}(s,-s_K)/\overline{W}_{+},
\ee
where
\begin{equation}
\Phi_{\pm}(s,s_1)=
Q^{\mu_{\pm}}_{\nu_{\pm}}(s)P^{\mu_{\pm}}_{\nu_{\pm}}(s_1)-
P^{\mu_{\pm}}_{\nu_{\pm}}(s)Q^{\mu_{\pm}}_{\nu_{\pm}}(s_1),
 \label{phix}
\end{equation}
\begin{equation}
\Psi_{\pm}(s,s_1)=
\left[Q^{\mu_{\pm}}_{\nu_{\pm}}(s)
\frac{{\rm d}}{{\rm d} s} P^{\mu_{\pm}}_{\nu_{\pm}}(s_1)-
P^{\mu_{\pm}}_{\nu_{\pm}}(s) \frac{{\rm d}}{{\rm d} s} 
Q^{\mu_{\pm}}_{\nu_{\pm}}(s_1)\right]
\;(s_1^2-1)
 \label{Mwyy2}
\end{equation}
for the case of low plus index (minus will be used below).
Here $Q^{\mu_{\pm}}_{\nu_{\pm}}(z)$ and $P^{\mu_{\pm}}_{\nu_{\pm}}(z)$
are the  same Legendre's functions, as in Sect.\ \ref{app1}, 
$s_K$ is given by 
(\ref{sk}), 
respectively. The constants
$\overline{W}_{\pm}$ independent of $s$ 
is related to the Wronskian (\ref{Wro}), (\ref{Wro2}) by 
\be
\overline{W}_{\pm}=(s^2-1)\; W_{\nu_{\pm}}^{\mu_{\pm}}(s)=
\frac{\Gamma\left(1+\nu_{\pm} + \mu_{\pm}\right)}
{\Gamma\left(1+\nu_{\pm}-\mu_{\pm}\right)} 
\label{Wpmt}
\ee
with
\begin{equation}
\mu_{\pm}  =i\sqrt{A_{\pm} +B_{\pm} },\qquad\qquad    \nu_{\pm} =
\frac{1}{2}(-1+i\sqrt{4A_{\pm} -1}),
 \label{munuindex}
\end{equation}
\begin{equation}
A_{+} =  \frac{12 \gamma\; (y_2-y_1)}{(y_3-y_1)}, \qquad\qquad
B_{+} = \frac{6(1 + 2 \gamma\;y_1)}{(y_3-y_1)}.
 \label{AB1plus}
\end{equation}
\begin{equation}
A_{-} = - \frac{12 (y_2-y_1)}{(y_3-y_1)}, \qquad\qquad
B_{-} = \frac{6(1 - 2 y_1)}{(y_3-y_1)}.
 \label{AB1minus}
\end{equation}
The solutions (\ref{y_Ath}) and (\ref{x_Ath}) are approximately periodic, 
the better the closer to the barrier energy. 
Note that in their derivations,
we found more conveniently to use the initial 
conditions at $t=T_A$ for R and $t=0$ for L orbits.
All energy-dependent quantities, $\mu_{\pm}$ and $\nu_{\pm}$ given in 
\eq{munuindex}, $\kappa$ and $a_\kappa$ in \eq{qmod}, as well as $s_K$ 
in \eq{sk}, are taken at the bifurcation energy $e=\est$ like $T_{\A}$ 
in this approximation.

For calculation of the $\trMRL$ \eq{tracR1} 
through the symmetry relations \eq{symrelmyyR} and \eq{symrelmyyL},
one has to derive the non-diagonal monodromy matrix elements 
$M_{yx}=\X_{yx}(T_A)$ and $M_{xy}=\X_{xy}(T_A)$. 
Neglecting the right-hand sides of the first and third equations 
in \eq{matrixstab0} and substituting their solutions 
\begin{equation}
\X^{(0)}_{xx}(s)=
\Psi_{+}(s,-s_K)/\overline{W}_{+}, \qquad\qquad
\X^{(0)}_{yy}(s)=\Psi_{-}(s,-s_K)/\overline{W}_{-},
\label{X0xxyys}
\end{equation}
into its second and forth equations, where the right-hand sides 
already contain small $v_x \propto \sqrt{e-e_n}$ 
\eq{vxV2} and $x_2 \propto \sqrt{e-e_n}$ of \eq{V1V2x2} near the
barrier, one finds for the solutions of the last two equations for 
$\X_{yx}(t)$ and $\X_{xy}(t)$, up to higher-order terms in the 
parameter $\sqrt{1-e_n}$,
\begin{equation}
\X_{yx}(t)=\frac{2\gamma}{a_{\kappa}\overline{W}_{-}}\int_0^t {\rm d} t_1
\,x_{po}(t_1)\,\Phi_{-}(s,s_1)\,\X^{(0)}_{xx}(s_1),
 \label{M1yx}
\end{equation}
\begin{equation}
\X_{xy}(t)=\frac{2\gamma}{a_{\kappa}\overline{W}_{+}}\int_0^t {\rm d} t_1
\,x_{po}(t_1)\Phi_{+}(s,s_1)\,\X^{(0)}_{yy}(s_1),
 \label{M1xy}
\end{equation}
see \eq{Mwyy2} for $\Psi_{\pm}(s,s_1)$ and \eq{phix} for $\Phi_{\pm}(s,s_1)$,
with the same relation of $t$ and $t_1$ to $s$ and $s_1$ through 
$s(t)$, see \eq{y_Ath}, as explained above. In these derivations, we used
the same transformation of equations of system \eq{matrixstab0}
to the Legendre form (\ref{legendre}) via $s=\tanh(z)$ like above.

With the help of \eq{M1yx} for $\M_{yx}=\X_{yx}(T_A)$, 
\eq{symrelmyyR} for $\M_{yy}$ of R, and \eq{M1xy} for $\M_{xy}=\X_{xy}(T_A)$, 
\eq{symrelmyyL} for $\M_{yy}$ of L orbits, and the periodic-orbit expressions 
(\ref{x_Ath}), by using the new variable $s(t)$ of \eq{y_Ath} in \eq{M1yx}
and \eq{M1xy}, one obtains 
\bea\l{trmmyx}
\trMR(e)&=&2 - 
\frac{4 \gamma\;(e-\est)}{3 a_{\kappa}^3\, V_2 
\overline{W}_{-} \overline{W}_{+}^2}
\int_{-s_K}^{s_K} \frac{\d s}{1-s^2}\,
\Phi_{+}\left(s_K,s\right)\, \Phi_{-}\left(s_K,s\right)\,
\Psi_{+}(s,-s_K),
\nonumber\\
\trML(e)&=&2 -
\frac{4 \gamma\;(e-\est)}{3 a_\kappa^2\, V_2 
\overline{W}_{-} \overline{W}_{+}^2}
\int_{-s_K}^{s_K} \frac{\d s}{1-s^2}\,
\Psi_{+}\left(s,-s_K\right)\, \Phi_{+}\left(s_K,s\right)\,
\Psi_{-}(s,-s_K).
\eea
All factors, except for $e-\est$, on right of 
\eq{trmmyx} can be considered at the bifurcation energy $e=\est$ for
$\est \to e \to 1$. We neglected here, as in previous subsections
of this Appendix, corrections of higher order in the small 
quantity $\sqrt{1-\est}$.

The integrals in \eq{trmmyx} can be taken analytically  
by simplifying their integrands with the approximation for the functions 
$\Phi_{-}(s_K,s)$ (\ref{phix}) and 
$\Psi_{-}(s,-s_K)$ (\ref{Mwyy2}) with indices ``$-$'', 
$\nu_{-}(e) \to -4$ and $\mu_{-}(e) \to -2$ valid well 
in the limit $e \to 1$, see
(\ref{munuindex}) and \cite{grry}, 
\be\l{phipsimin}
\Phi_{-}(s_K,s)  \approx 
\frac{\overline{\Gamma}\; \overline{W}_{-}}{60 \left(1-s_{K}^2\right)}
\,P_{3}^{-2}(s),\qquad\qquad 
\Psi_{-}(s,-s_K) \approx  
-\frac{s_K \overline{\Gamma}\; \overline{W}_{-}}{30 \left(1-s_K^2\right)}
\,P_{3}^{-2}(s), 
\ee
where the indices ``$-$'' appear only through a smooth energy-dependent 
coefficient,
\be\l{gammabar}
\overline{\Gamma}=\frac{\Gamma\left(-\mu_{-}-\nu_{-}-1\right)
\left(\nu_{-}+\mu_{-}+1\right)}
{\Gamma\left(\mu_{-}-\nu_{-}-1\right)\left(\nu_{-}-\mu_{-}+1\right)},
\ee
and $\overline{\Gamma}(e) $ tends to 120 smoothly at 
$e \rightarrow 1$, see \eq{gammabar} and \eq{munuindex}. 
Notice that the contribution of the correction \cite{grry} 
to this approximation is negligibly small for the calculation 
of this integral, being of relative order $\sqrt{1-\est}$. 
Therefore, within the approximation (\ref{phipsimin}), 
the integrals over $s$ in \eq{trmmyx} are reduced to the sum of several
standard indefinite integrals of the products of two Legendre's 
functions with weight $s$ and 
indices $\nu=\nu_{+}$ and $\mu=\mu_{+}$ of (\ref{munuindex})  
\cite{prud},
\bea\l{intleg}
\int \d s\, s {\cal L}_{\nu}^{\mu}(s)\,\overline{{\cal L}}_{\nu}^{\mu}(s) &=&
{\cal R}_{\nu \mu}^{{\cal L}\overline{{\cal L}}}(s)=\aleph_1 
{\cal L}_{\nu}^{\mu}(s)\, {\overline {\cal L}}_{\nu}^{\mu}(s) - \aleph_2 
\left[{\cal L}_{\nu}^{\mu}(s)\, \overline{{\cal L}}_{\nu+1}^{\mu}(s) +
\overline{{\cal L}}_{\nu}^{\mu}(s)\, {\cal L}_{\nu+1}^{\mu}(s)\right] 
\nonumber\\
&-&\aleph_3 \overline{{\cal L}}_{\nu+1}^{\mu}(s)\, {\cal L}_{\nu+1}^{\mu}(s), 
\eea
where ${\cal L}_{\nu}^{\mu}$ and 
$\overline{{\cal L}}_{\nu}^{\mu}$ are any pair of the Legendre
functions from the set $P_{\nu}^{\mu}$, $Q_{\nu}^{\mu}$,
\be\l{constN}
\aleph_1=\frac{\mu^2 - (\nu+1)(\nu+s^2)}{2 \nu (\nu+1)}, \qquad 
\aleph_2= \frac{(\nu+1)(\mu-\nu-1) s}{2 \nu (\nu+1)},\qquad 
\aleph_3=\frac{\left(\mu-\nu-1\right)^2}{2 \nu (\nu+1)}.
\ee 
The strong energy dependence of $\trM(e)$ (\ref{tracR1}) 
is coming through the $\overline{W}_{-}(e)$ or $1-s_K^2(e)$  
which tend both to zero for $e \to 1$ via the approximate key relations  
\be\l{Wminuse}
\overline{W}_{-}(e) \approx  -\frac{35 \sqrt{3 (1-e)}}{441\; 
\overline{\Gamma}}, 
\qquad\qquad
1-s_K^2(e) \approx \sqrt{\frac{1-e}{27}},
\ee
see
\eq{Wpmt} for $\overline{W}_{-}$, \eq{munuindex} for $\mu_{-}$
and $\nu_{-}$ \cite{grry}. 
The key point
of our transformations is to remove indeterminacy zero by zero 
by identical cancellation of the singular factor
$\overline{W}_{-}(\est)$ near the saddle from the 
denominators  
and that of the functions (\ref{phipsimin}) 
with indices ``$-$'' in the numerators of the integrands. 
Then, another constant 
singular factor $1-s_K^2$ can be taken off the integrals. Thus, 
after such simple algebraic transformations with help of 
\eq{phipsimin}
and \eq{intleg}, from \eq{trmmyx} 
one obtains 
\bea\l{trmmyxan}
&&\trMRL(e)=2 \pm \gamma\;\zeta_{RL}(\est)\,\left(e-\est\right)\; 
 \left(1-s_K^2\right)\left\{{\cal D}_{\nu \mu}^{PP}(s_K)
Q_{\nu}^{\mu}(s_K) 
\frac{\d}{\d s}Q_{\nu}^{\mu}(-s_K)
+ {\cal D}_{\nu \mu}^{QQ}(s_K)\right.
\nonumber\\
&&\times \left.P_{\nu}^{\mu}(s_K)\;\frac{\d}{\d s}P_{\nu}^{\mu}(-s_K)
-{\cal D}_{\nu \mu}^{PQ}(s_K)\left[
Q_{\nu}^{\mu}(s_K) 
\frac{\d}{\d s}P_{\nu}^{\mu}(-s_K)+P_{\nu}^{\mu}(s_K) 
\frac{\d}{\d s}Q_{\nu}^{\mu}(-s_K)\right]\right\},  
\eea
where
\be\l{zetaRL}
\zeta_{RL}(e)=\sqrt{\frac{8 s_K\; b_1(e)}{3\overline {W}_{+} V_2^2(e)}}
\left\{{1/s_K \qquad \rm{for}\qquad \R 
\atop{2 \it{a}_\kappa \qquad \,\,\,\rm{for}\qquad\, \L}}\right\},
\ee
\begin{equation}
b_1(e)=\frac{1}{6a_\kappa^6 \overline{W}_{+}^3}\;
\left(\frac{\overline{\Gamma}}{120}\right)^2\;
\frac{s_K}{\left(1-s_K^2\right)^2} \approx
\frac{288}{\overline{W}_{+}^3(1-e)},
 \label{b1}
\end{equation}
\be\l{Ds}
{\cal D}_{\nu \mu}^{{\cal L} \overline{\cal L}}(s)= 
{\cal R}_{\nu \mu}^{{\cal L} \overline{\cal L}}(s)-
{\cal R}_{\nu \mu}^{{\cal L} \overline{\cal L}}(-s_K),
\ee
at $s=s_K$
with the indices ``+'',  regular in the considered limit, 
$\mu=\mu_{+}$ and $\nu=\nu_{+}$.
Note that the derivatives of the Legendre functions on the
right of \eq{trmmyxan} are approximately proportional to $ 1/(1-s_K^2)$,
according to the recurrence relations for the Legendre functions with 
indices ``+'' of \eq{munuindex}
\cite{grry}, and therefore, the product of the factor $~1-s_K^2~$ by 
the expression in figure brackets is a smooth function 
of the energy $\est$ near the saddle as well as $\overline{W}_{+}$,
see \eq{Wpmt} and \eq{munuindex}. Therefore, the strongest energy 
dependence near the saddle 
is coming only from the coefficient $\zeta_{RL}(e)$ (\ref{zetaRL}).
By making use of asymptotic expressions (\ref{b1}) for $b_1(e)$ and 
(\ref{vxV2}) for $V_2(e)$ through \eq{zetaRL} for $\zeta_{RL}$ 
in the limit $\est \to 1$, up to higher order terms in
small parameter $\sqrt{1-\est}$, from \eq{trmmyxan} we arrive 
at the result
\bea\l{trmmyxanas}
\trMRL(e)&=&2 \pm \gamma\,\frac{24 (e-\est)}{1-\est}\;
\frac{1-s_K^2}{\overline {W}_{+}^2} 
\left\{\!{\cal D}_{\nu \mu}^{PP}(s_K)
Q_{\nu}^{\mu}(s_K) 
\frac{\d}{\d s}Q_{\nu}^{\mu}(-s_K)+
{\cal D}_{\nu \mu}^{QQ}(s_K) P_{\nu}^{\mu}(s_K)\right.~
\nonumber\\
&\times& \left.
\!\!\!\!\frac{\d}{\d s}P_{\nu}^{\mu}(-s_K)\!
-\!{\cal D}_{\nu \mu}^{PQ}(s_K)\!\!\left[
Q_{\nu}^{\mu}(s_K) 
\frac{\d}{\d s}P_{\nu}^{\mu}(-s_K)+P_{\nu}^{\mu}(s_K) 
\frac{\d}{\d s}Q_{\nu}^{\mu}(-s_K)\right]\!\right\} \!.  
\eea

Using the asymptotic forms of the Legendre functions in
figure brackets through hypergeometric series (cf.\ \cite{grry}, 
Eqs.\ 8.704, 8.705, and 8.737), like for the derivation of $\trMA(e)$ 
(\ref{trMTA3}), we may expand the function $s_K$ in \eq{trmmyxanas} 
in the small parameter $1-s_K(\est) \propto \sqrt{1-\est}$, see 
\eq{Wminuse}, in the limit $e_n \to 1$. Up to higher terms of
relative order $\sqrt{1-\est}$, we then obtain the asymptotic 
result for $\trMRL^{(as)}(e)$ given in (\ref{trMTRLas}), 
correctly describing the ``GHH fans'', with the slope function 
$c_{RL}(\gamma)$ given in \eq{crl}.

\section{Perturbative calculation of $\trMRL(e)$ near $e=1$}
\l{appc2}

\subsection{``Frozen approximation'' (FA) for the periodic orbits}
\l{appc2subb}

Within the FA, we set $y_{po}(t) \approx y_{A}(t)$ (cf.\ \sec{sectrmrl}). 
In order to find $\M_{yy}$ of \eq{tracR1} for $\trM$, 
we solve the system of equations (\ref{matrixstab0})
for $\X_{yy}(t)$ and $\X_{xy}(t)$, 
with the initial conditions \eq{initialcond},
iteratively by exploiting the smallness of their r.h.\ sides.
Substituting expansions \eq{xyexp} and \eq{Xyyxyexp} into 
these equations at zero and first order in $\expeps$, 
respectively, and then solving them for the monodromy 
matrix element $\M_{yy}=\X_{yy}(T_{\A})$, one obtains 
\begin{equation}
\M_{yy}=1 + (e-\est)\;\M_{yy,1}^{(1)}(\est).
 \label{MyyTA}
\end{equation}
The first term is given by $M_{yy}(\est)=1$ at order zero of the
perturbation scheme, see \eq{myydxdyR}. For the coefficient 
$\M_{yy,1}^{(1)}(\est)$ in the linear term of \eq{MyyTA}
for the R type orbits, one finds  
\begin{equation}
\M_{yy,1}^{(1)}(\est)=-\gamma^2 b_1(\est)\;
I_{yy,1}^{(1)}(\est),
 \label{Iyyfin}
\end{equation}
where $b_1(e)$ is given by \eq{b1}, 
\bea\l{Iyy0sfin}
I_{yy,1}^{(1)}(\est)&=&\int_{-s_K}^{s_K} {\rm d}s\;s\;
\Phi_{+}(s_K,s)
\left\{Q_{\nu}^{\mu}(s_K)\left[Q_{\nu}^{\mu}(s)
{\cal D}_{\nu \mu}^{PP}(s) - P_{\nu}^{\mu}(s)
{\cal D}_{\nu \mu}^{PQ}(s)\right]\right.
\nonumber\\
&-& \left. P_{\nu}^{\mu}(s_K)\left[Q_{\nu}^{\mu}(s)
{\cal D}_{\nu \mu}^{PQ}(s) - P_{\nu}^{\mu}(s)\;
{\cal D}_{\nu \mu}^{QQ}(s)\right]\right\},
\eea
${\cal D}_{\nu \mu}^{{\cal L} \overline{{\cal L}}}$ is 
the matrix \eq{Ds}. 
For L orbits, one has similar derivations.
All quantities on the r.h.s.\ of \eq{Iyy0sfin} are 
taken at the bifurcation energy $e=\est$. In these derivations,
the double integrals were reduced to single integrals
through simple algebraic transformations with the help of 
\eq{phipsimin} and \eq{intleg}, canceling the singular multiplier
$\overline{W}_{-}$, see \eq{Wminuse}, from the denominator with that  
in the numerator functions (\ref{phipsimin}) in the integrand  
near the saddle, like in Appendix \ref{app2subb}.

\subsection{Corrections to the FA}
\l{appc2subc}

The results \eq{MyyTA}-\eq{Iyy0sfin} 
can be improved much beyond the FA by taking into account 
the next-order terms in the expansion \eq{xyexp}  of $y_{po}(t)$.
We then find more exact solutions to \eq{motiony}. For instance, 
for the R orbits we get $x_{\R}(t)=u_{\R}\;x_{\R}^{(0)}(t)$ 
and $y_{\R}(t)=y_{\A}(t) + \expeps \;y_{\R}^{(1)}(t)$, 
which obey the initial conditions
$y_R^{(1)}(0)=0$ and $\dot{y}_R^{(1)}(0)=0$, whereby 
\begin{equation}
y_{\R}^{(1)}(t)=
\frac{\gamma}{3 a_\kappa^2 \overline{W}_{-}}\;
\int_{-s_K}^{s} \frac{\d s_1}{1-s_1^2} \left[x_{\R}^{(0)}(t_1)\right]^2 
\Phi_{-}(s_1,s)
 \label{y2eps}
\end{equation}
with the relations $s=s(t)$ and $s_1=s(t_1)$ of \eq{y_Ath}.
By making use of this solution and the corresponding more exact
expansion \eq{Xyyxyexp} for $\X_{yy}(t)$ of the problem 
\eq{matrixstab0} and \eq{initialcond}, one finds correction to 
$\trMR(e)$. For these more exact calculations, we have to extend 
\eq{MyyTA} to the complete solution for $\M_{yy}$, collecting 
all leading corrections of first order in $\expeps$, 
\begin{equation}
\M_{yy}= 
1+ (e-\est)\left[\M_{yy,1}^{(1)}(\est)+
\M_{yy,2}^{(1)}(\est)\right],
 \label{MyyTAeps}
\end{equation}
where
\begin{equation}
\M_{yy,2}^{(1)}(\est)=\frac{2}{a_\kappa \overline{W}_{-}} 
\int_{0}^{T_{\A}} {\rm d}t_2\,
\X^{(0)}_{yy}(s_2)\,\Phi_{-}(s_K,s_2)\,
y_R^{(1)}(t_2)
 \label{M02epsint1}
\end{equation}
with \eq{y2eps} for $y_R^{(1)}(t_2)$.
After a change of the integration variable from $t_2$ to
$s_2=s(t_2)$ of \eq{y_Ath} in (\ref{M02epsint1}) and
using expression (\ref{y2eps}) for $y_R^{(1)}(t_2)$,
we may use the same approximations (\ref{phipsimin}), with the
help of \eq{X0xxyys} for $\X^{(0)}_{yy}(s_2)$, to 
perform analytically the integral in \eq{y2eps} in terms of elementary 
functions. 
Finally, after canceling identically the singular factor 
$\overline{W}_{-}$ and another singular factor $1-s_K^2$ from
the remaining integral, like in Appendix \ref{app2subb}, see \eq{Wminuse},
one arrives at 
\be\l{M02epsintTAfin}
\M_{yy,2}^{(1)}(\est)=-\gamma \;b_2(\est)\;I_{yy,2}^{(1)}(\est),
\qquad\qquad b_2(\est)=\frac{1}{2}\overline{\Gamma}\;\overline{W}_{+}b_1(\est),
\ee
where 
\bea\l{Iyy02sfin}
I_{yy,2}^{(1)}(\est)&=&
\int_{-s_K}^{s_K} {\rm d} s
\frac{\Phi_{+}^2(s_K,s)}{1-s^2}\; 
\left\{P_{3}^{-2}(s) \left[{\cal F}_{Q}(s)-{\cal F}_{Q}(s_K)\right]\right. 
\nonumber\\
&-&\left.
 Q_{3}^{-2}(s) \left[{\cal F}_{P}(s)-{\cal F}_{P}(s_K)\right]
\right\},
\eea
\bea\l{funfs}
{\cal F}_{Q}(s)&=& 
\frac{1}{5760}\left\{30 s -118 s^3 +210 s^5 -90 s^7 
-15 \left[{\rm ln}\left(\frac{1+s}{1-s}\right)\right.\right. 
\nonumber\\
&+&\left.\left.
192 s^4 {\cal F}_{P}(s)\; {\rm ln}\left(\frac{1+s}{1-s}\right)\right]\right\},
\qquad {\cal F}_{P}(s)=
\frac{1}{192}\left(6 s^4 - 8 s^6 + 3 s^8\right).
\eea

Taking into account both energy corrections in \eq{MyyTAeps} 
with \eq{Iyyfin} and \eq{M02epsintTAfin}, we transform \eq{MyyTAeps} for
$\trMR$ into the asymptotic result
\be\l{trMR2eps}
\trMR(e)  
= 2 - 2 (e-\est)\; \left[\gamma^2\;b_1(\est) I_{yy,1}^{(1)}(\est) +
\gamma b_2(\est) I_{yy,2}^{(1)}(\est)\right]
+ {\cal O}\left[\frac{(e-\est)^2}{(1-\est)^{1/2}}\right].
\ee
Similar expression for the stability trace $\trML(e)$ can easy obtained for
the L orbits. As seen from \eq{M02epsintTAfin}
and \eq{b1}, $b_2(\est) \propto b_1(\est) \propto 1/(1-\est)$, 
and the other factors $I_{yy,1}^{(1)}(\est)$ \eq{Iyyfin} and
$I_{yy,2}^{(1)}(\est)$ \eq{Iyy02sfin} are smooth functions
of $\est$, as confirmed by numerical integrations in \eq{Iyyfin}
and \eq{Iyy02sfin}. Therefore, both corrections in \eq{trMR2eps}
are  mainly proportional to $(e-\est)/(1-\est)$, i.e., 
linear in the $(e-\est)$. They are both finite in the barrier limit $e \to 1$
but numerically, the essential contribution to (\ref{trMR2eps}) is coming 
from the first FA correction while the second one (above FA) is
much smaller. Note that the leading higher-order terms in the parameter
$\epsilon$ (\ref{expeps}), originating from the next 
iterations in the perturbation scheme \eq{xyexp} and \eq{Xyyxyexp}, can be 
estimated, in fact, as of higher order in $\sqrt{1-\est}$. 
Thus, the complete sum of energy-dependent corrections 
in \eq{trMR2eps} has the same leading energy dependence
$\propto (e-\est)/(1-\est)$,
up to higher-order terms in small parameter $\sqrt{1-\est}$, 
as in \eq{trMTRLas}.
The leading energy dependence of $\trM(e)$ 
is thus precisely that found explicitly in the non-perturbative result 
(\ref{trmmyxanas}) in Appendix \ref{app2subb}.

\begin{figure}
\begin{center}
\epsfig{figure = 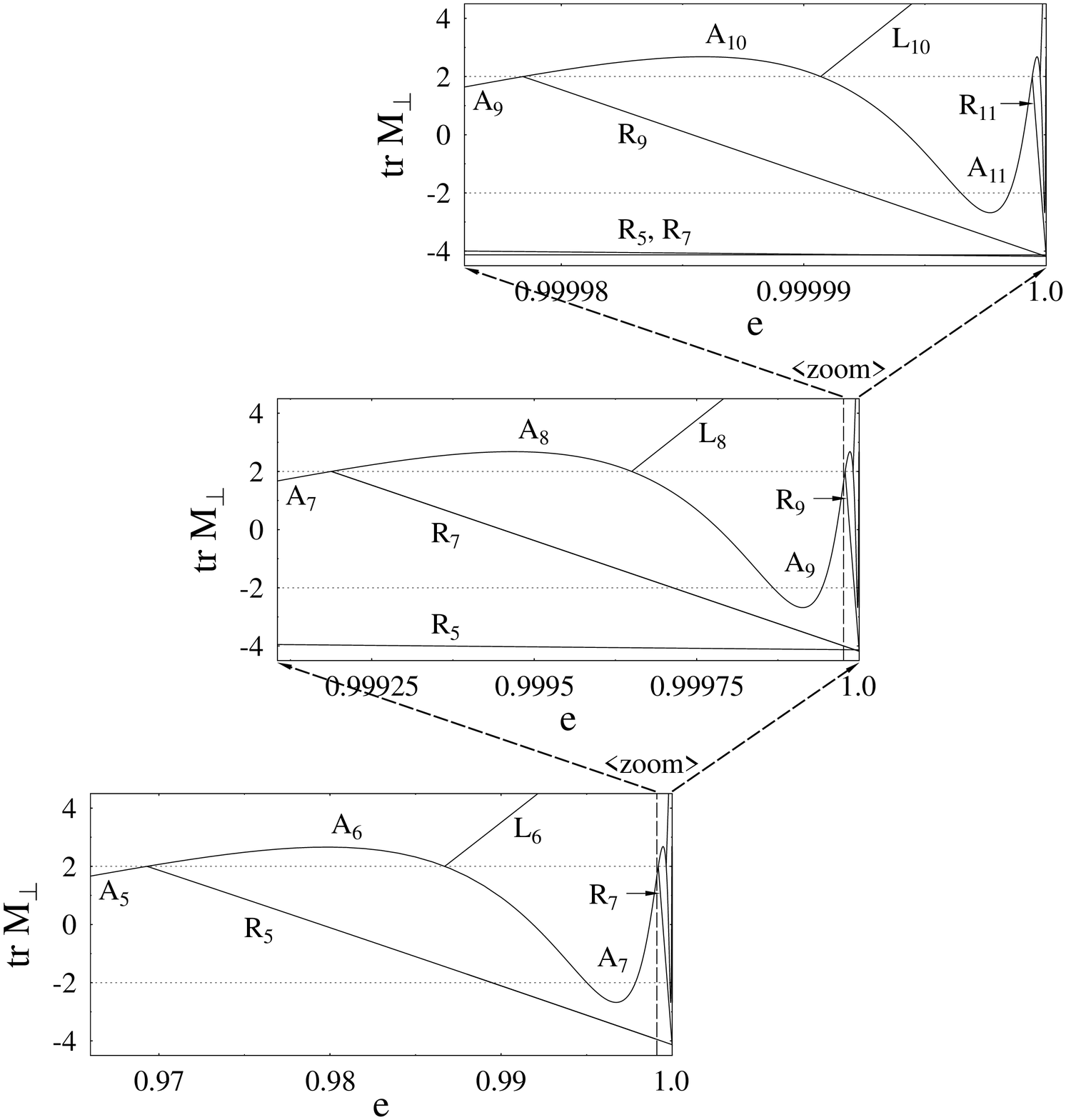,width = 0.9\textwidth,clip}
\end{center}
\caption{
Trace of stability matrix $\Mt$ of orbit A and the orbits born at
successive pitchfork bifurcations in the standard HH system ($\gamma=1$), 
plotted versus the scaled energy $e$. 
{\it From bottom to top:} successively zoomed energy scale near $e=1$ 
(from \cite{mbgu}).\label{zoom}
}
\end{figure}

\begin{figure}
\begin{center}
\epsfig{figure = 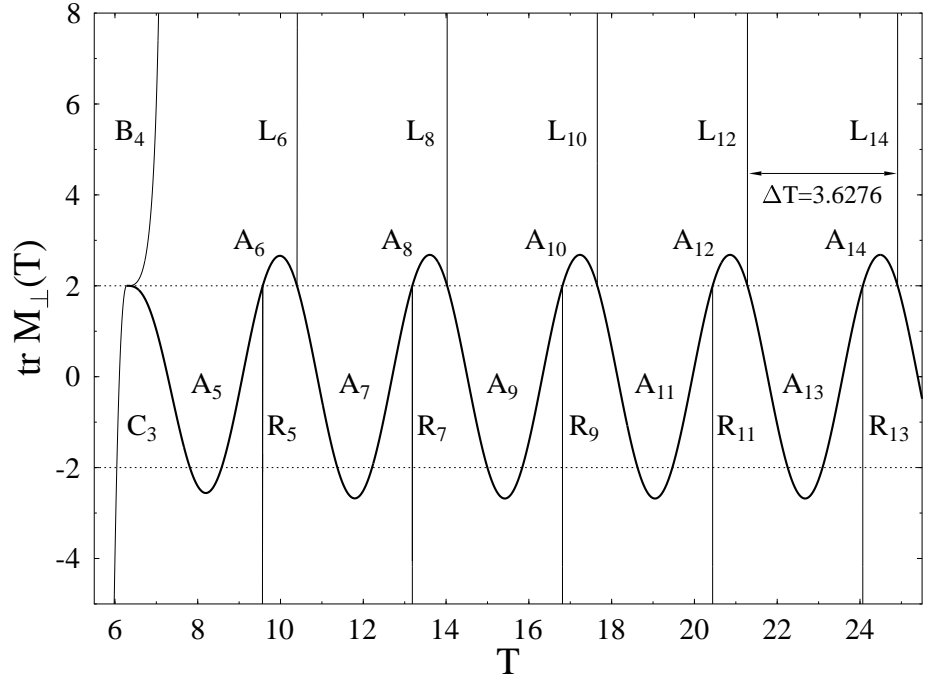,width = 0.9\textwidth,clip}
\end{center}
\caption{
Trace of the stability matrix $\Mt$ of the orbits A (heavy line), B, C, and 
the orbits R$_{2m-1}$, L$_{2m}$ ($m\geq3$) born at successive pitchfork
bifurcations of orbit A in the standard HH potential,
plotted versus their individual periods $T$. $\Delta T$ is the 
asymptotic period of the curve $\trMA\,(T_A)$ for large $T_A$ 
(from \cite{mbgu}).
}\label{hhtrmt}
\end{figure}

\begin{figure}
\begin{center}
\epsfig{figure = 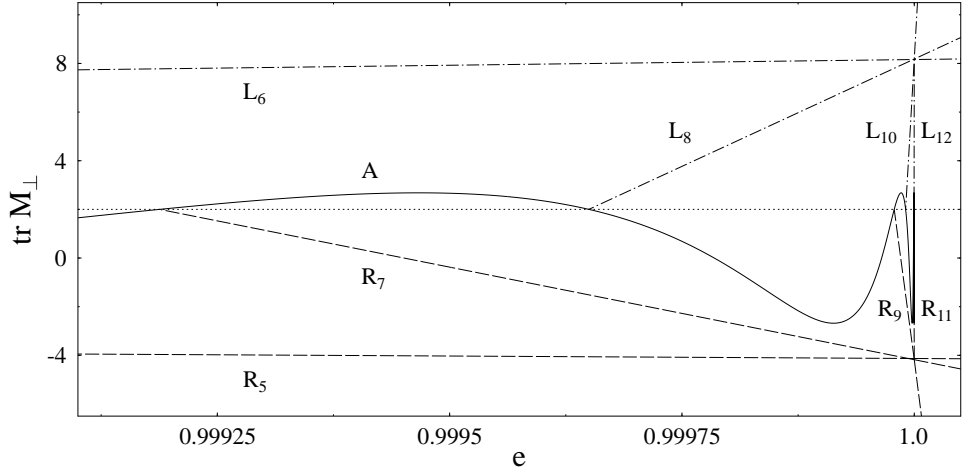,width = 0.9\textwidth,clip}
\end{center}
\caption{
The ``H\'enon-Heiles fans''. Trace of stability matrix of primitive A orbit
(solid line) and the first four pairs of R orbits (dashed) and L orbits
(dash-dotted) in the standard HH system, plotted versus scaled energy
$e$; the latter forming two fans for the R and L orbits.
} \label{hhfane}
\end{figure}

\begin{figure}
\begin{center}
\epsfig{figure = 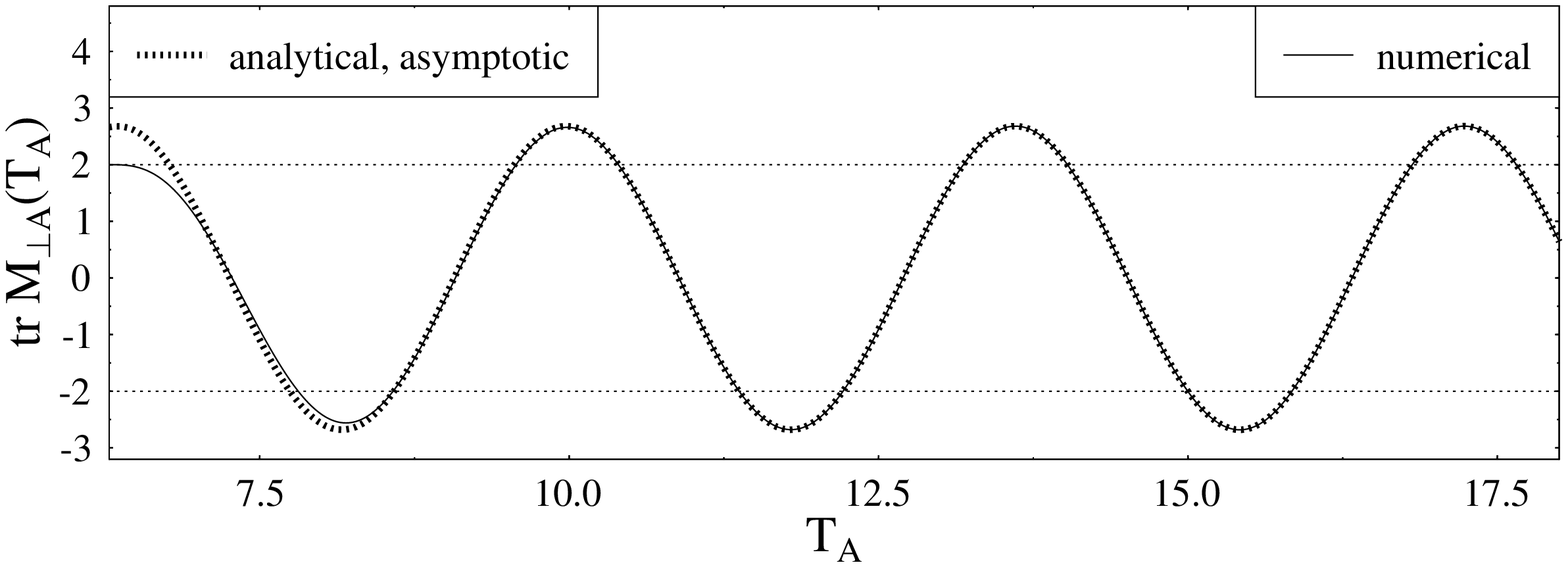,width = 0.9\textwidth,clip}
\end{center}
\caption{
Stability discriminant $\trMA$  of the A orbit in the HH potential
($\gamma=1$), plotted versus period $\TA$. {\it Solid line:}
numerical result (as in \fig{hhtrmt}, from \cite{mbgu}). 
{\it Dotted line:} analytical asymptotic result $\trMA^{(as)}\,(T_A,1)$ 
given in \eq{trMTHH}.
}\label{hhtrmas}
\end{figure}

\begin{figure}
\begin{center}
\epsfig{figure = 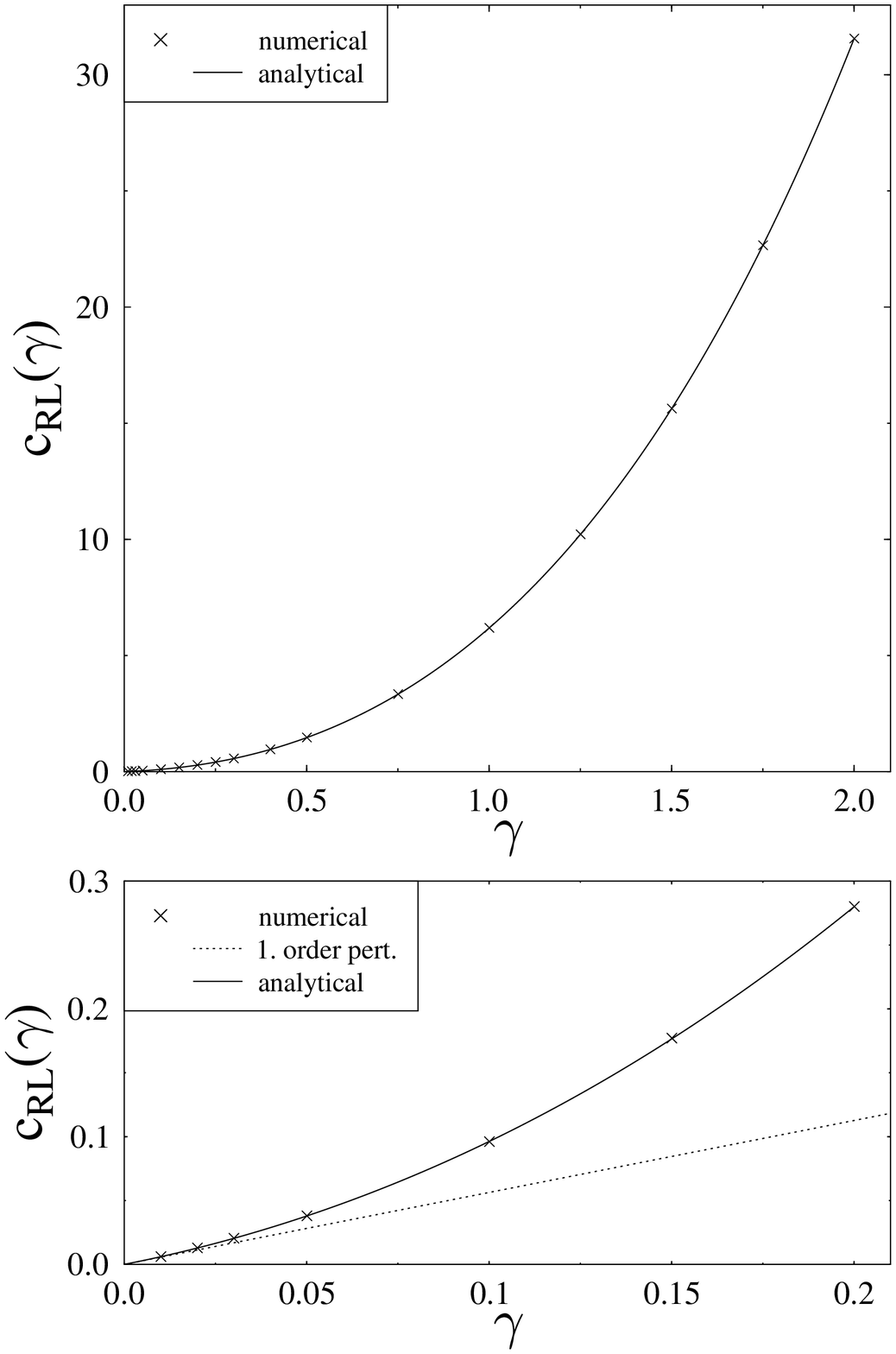,width = 0.9\textwidth,clip}
\end{center}
\caption{
{\it Upper panel:} slope parameter $c_{RL}$ of the ``HH fans'' 
plotted versus the potential parameter $\gamma$. 
{\it Crosses:} numerical values; {\it solid
line:} the function $c_{RL}(\gamma)$ given in \eq{crl}.
{\it Lower panel:} excerpt for small values of $\gamma$. The
dotted line gives the linear approximation to  $c_{RL}(\gamma)$,
with the slope given in \eq{crlp}, as found in a semiclassical
perturbative approach \cite{bkwf}.
}\label{fanslope}
\end{figure}

\begin{figure}
\begin{center}
\epsfig{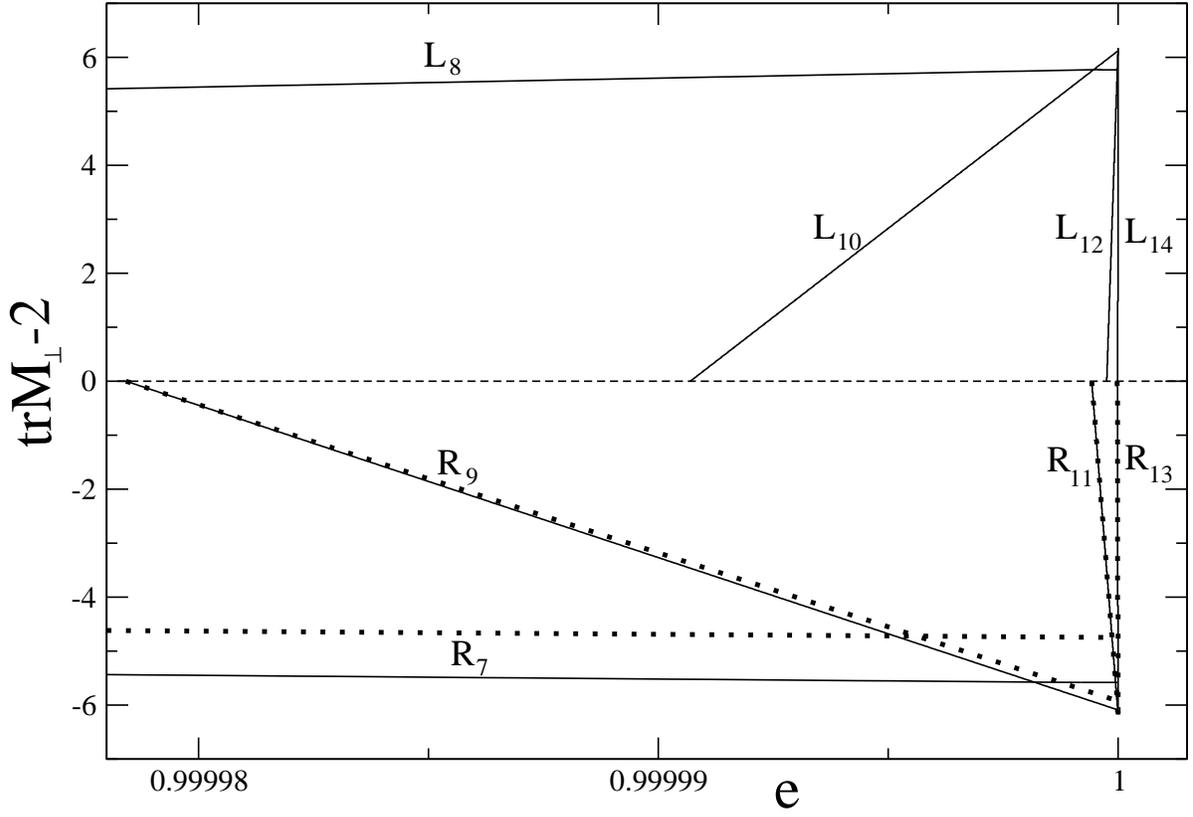}
\end{center}
\caption{
Stability traces $\,\trMRL-2\,$
as functions of the energy $e$  
at $\gamma=1$. Solid lines show the analytical expression 
(\ref{trmmyxan}) for $\R_{n}$ and $\L_{n}$ orbits with $n=7-14$. 
Dashed lines are the perturbative results (\ref{trMR2eps}) for a 
few R orbits as examples.
}\label{fig1D}
\end{figure}

\begin{figure}
\begin{center}
\epsfig{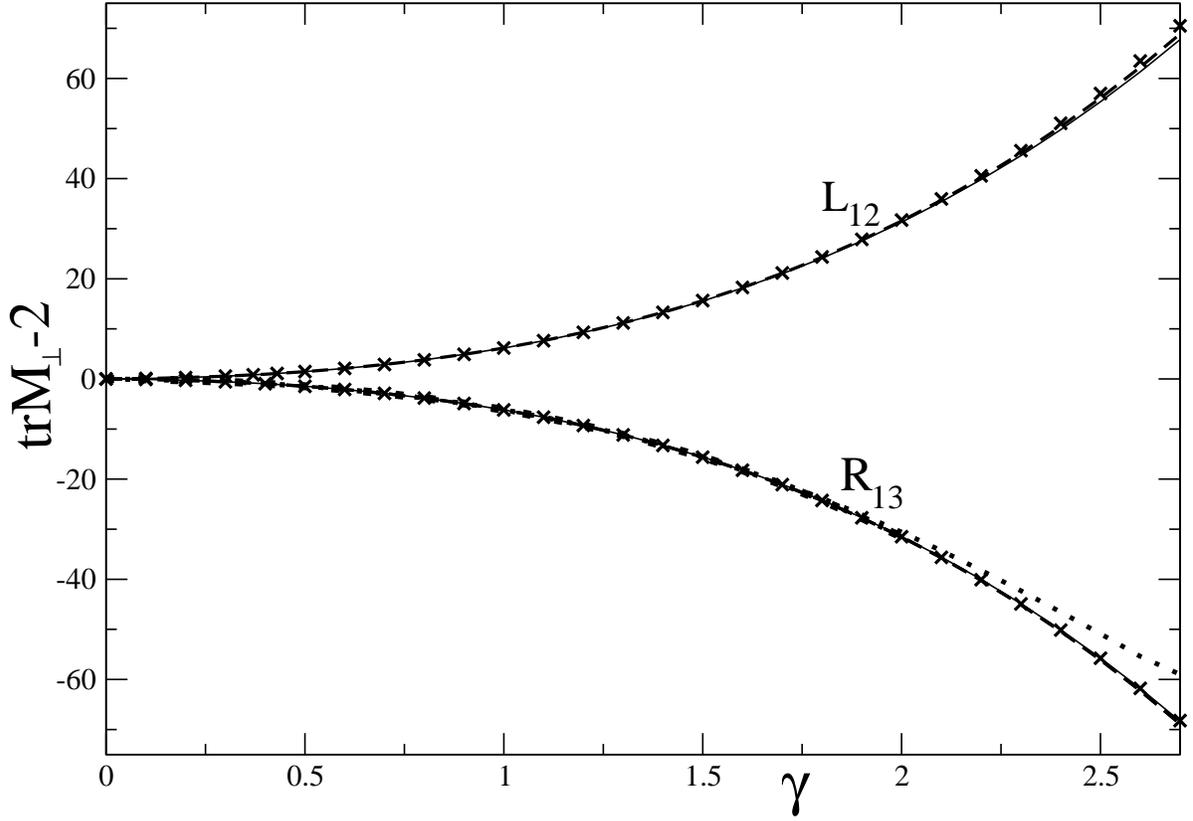}
\end{center}
\caption{
Stability traces $\trMRL-2$ as functions of $\gamma$ for R and L
orbits, respectively, evaluated at the barrier energy $e=1$.
Solid lines show the analytical expression (\ref{trmmyxan})  
for the orbits $\R_{13}$ and $\L_{12}$; dashed lines the 
asymptotic results (\ref{crl}) for $\mp c_{RL}(\gamma)$;
dots are the perturbative results to \eq{trMR2eps};
and crosses show the numerical results $\mp d_n^{num*}$ with the 
FA initial conditions as in \tab{table1RL}. The bifurcation
energies $e_n(\gamma)$ are obtained analytically through Eqs.\ 
(\ref{enas}).} \label{figghhan}
\end{figure}

\Table{tabenas}{16.6}{
\begin{tabular}{|c|l|l|}
\hline
$n$ & ~~~~~~~~~~~$e_n^*$ & ~~~~~~~~$e_n$ \\
\hline
5  & 0.96945 52246 81049 & 0.96930 90904 \\
6  & 0.98682 99363 40510 & 0.98670 92353 \\
7  & 0.99918 81219 03970 & 0.99918 78410 \\
8  & 0.99964 99405 84051 & 0.99964 98    \\
9  & 0.99997 84203 34217 & 0.99997 8390  \\
10 & 0.99999 06954 44011 & 0.99999 06955 \\
11 & 0.99999 94264 13919 & 0.99999 9424  \\
12 & 0.99999 97526 85521 & 0.99999 97525 \\
13 & 0.99999 99847 54120 & 0.99999 99847 5   \\
14 & 0.99999 99934 26398 & 0.99999 99934 3   \\
15 & 0.99999 99995 94766 & 0.99999 99996 046 \\
16 & 0.99999 99998 25274 & 0.99999 99998 249 \\
\hline
\end{tabular}
\vspace*{-0.3cm}
}{~~~Bifurcation energies in the standard HH potential ($\gamma=1$).
$e_n^*$: asymptotic values, calculated from the analytical 
expressions \eq{enas} up to 15 digits with MATHEMATICA.
$e_n$: numerical values taken from \cite{lamp}.}

\Table{table1RL}{16.6}{
\begin{tabular}{|c|c|c|c|c||c|c|c|c|}
\hline
$n$ & $d_{n}^{sa}$ &
 $d_{n}^{an}$ & 
 $d_{n}^{num*}$ & $d_{n}^{num}$ &
 $n$ & $d_{n}^{an}$ &
 $d_{n}^{num*}$ & $d_{n}^{num}$\\
\hline
7  & 4.7476&   5.5863&   6.1688 & 6.1801&
8 & 5.7796 & 6.2661 &6.1803\\
9 & 5.9234&    6.0901 &  6.1800 & 6.1819& 
10 & 6.1209 & 6.1951 &6.1820\\
11 & 6.1391&   6.1685 &  6.1817 & 6.1820& 
12 & 6.1731 &  6.1841 &6.1897\\
13 & 6.1750&   6.1801 &  6.1819 & 6.1837& 
14 &  6.1807 & 6.1823 &\\ 
15 & 6.1808&   6.1817 &  6.1820 &&
16 &6.1818 & 6.1821 &\\ 
17 & 6.1818&   6.1820 &  6.1820 && 
18 &  6.1820 &  6.1820 &\\
19 & 6.1820&   6.1820 &  6.1820 &&
20 & 6.1820 & 6.1820 &\\
\hline
\end{tabular}
}{~~~The 
slope parameters 
$d_n^{sa}$ of the semi-analytical \eq{trMR2eps}  
and $d_n^{an}$ of the analytical \eq{trmmyxan} expressions
vs the numerical values
$d_{n}^{num*}$ for solving \eq{motionx}, \eq{motiony},
\eq{Xoft} within the FA for the initial 
conditions at the top point (\ref{toppoint}) 
of the periodic $R_n$ (left) and $L_n$ (right) orbits,
and $d_{n}^{num}$ is exactly full numerical results \cite{lamp}
($\gamma=1$ in all cases).}

\end{document}